\date{}
\documentclass[aps,pra,showpacs]{revtex4}
\usepackage{graphicx,amsmath,amssymb,amsfonts,latexsym,color,dcolumn}

\begin{document}

\title{Quantum Vacuum Fluctuations in Presence of Dissipative Bodies: Dynamical Approach for Non-Equilibrium and Squeezed States}

\author{Adri\'an E. Rubio L\'opez\footnote{adrianrubiolopez0102@gmail.com}}

\affiliation{Departamento de F\'\i sica {\it Juan Jos\'e Giambiagi}, FCEyN UBA and IFIBA CONICET-UBA, Facultad de Ciencias Exactas y Naturales, Ciudad Universitaria, Pabell\' on I, 1428 Buenos Aires, Argentina}

\date{today}

\begin{abstract}
The present work contributes to the study of non-equilibrium aspects of the Casimir forces with the introduction of squeezed states in the calculations. Throughout this article two main results can be found, being both strongly correlated. Primarily, the more formal result involves the development of a first-principles canonical quantization formalism to study the quantum vacuum in presence different dissipative material bodies in completely general scenarios. For this purpose, we considered a one-dimensional quantum scalar field interacting with the volume elements' degrees of freedom of the material bodies, which are modeled as a set of composite systems consisting in a quantum harmonic oscillators interacting with an environment (provided as an infinite set of quantum harmonic oscillators acting as a thermal bath). Solving the full dynamics of the composite system through its Heisenberg equations, we studied each contribution to the field operator by employing general properties of the Green function. We deduced the long-time limit of the contributions to the field operator. In agreement with previous works, we showed that the expectation values of the components of the energy-momentum tensor present two contributions, one associated to the thermal baths and the other one associated to the field's initial conditions. This allows us to the direct study of steady situations involving different initial states for the field (keeping arbitrary thermal states for the baths). This leads to the other main result, consisting in computing the Casimir force when the field is initially in thermal or continuum-single-mode squeezed states (being the latter characterized by a given bandwidth and frequency). Time-averaging is required for the squeezed case, showing that both results can be given in a unified way, while for the thermal state, all the well-known equilibrium results can be successfully reproduced. Finally, we compared the initial conditions' contribution and the total force for each case, showing that the latter can be tuned in a wide range of values through varying the size of the bandwidth.
\end{abstract}


\maketitle

\section{Introduction}

Quantum nature of vacuum is still one of the most relevant features of quantum field theory (QFT) due to theoretical and also technological implications. From a conceptual point of view, understanding the physical properties of vacuum at the quantum level is unavoidable in a wide range of areas, from quantum optics\cite{BarnettRadmore} and condensed matter\cite{FradkinGitmanShvartsman} to astrophysics and cosmology\cite{Calzetta1987}. On the other hand, from a practical point of view, novel experiments measuring forces\cite{BimonteLopezDecca,MundayCapassoParsegian} and heat transfer at the nanoscale\cite{Kittel} shows that, exploiting all these features, a new generation of technological improvements is coming\cite{SerryWalliserMaclay,BuksRoukes}.

For these reasons, study quantum vacuum fluctuations (going from Casimir-Van der Waals interactions to quantum friction) is of main interest. In particular, Casimir forces arising from the physical adaptation of a quantum field to the presence of arbitrary-shaped objects acting as boundaries, are still a source of abundant scientific research. Since the foundational paper by Casimir\cite{Casimir1948}, where the force between two perfect conductor plates was studied, numerous subsequent works pointed in the direction to study different aspects of these vacuum phenomena more related to realistic experimental scenarios. The natural step and one of the most remarkable works in this sense, is the work by Lifshitz\cite{Lifshitz1956}, where dissipation and noise were included for the first time in the calculation of the Casimir force between dielectric plates at zero temperature through an approach based on stochastic electrodynamics, resulting in the celebrated Lifshitz formula for the force. This work also set the basis for the so-called fluctuational quantum electrodynamics (FQED) as it is currently known. Few years later, in Ref.\cite{DzyLifPita}, a formal approach based on QFT at finite temperature was developed to study thermal corrections on computing the force, arriving to a more sophisticated and complete form of the Lifshitz formula at thermal equilibrium. After these works, the study of different features of the Casimir effect had a significant growth going beyond fundamental physics and entering chemistry and biology\cite{KardarGolestanian}. Moreover, a new generation of extremely accurate experiments (see for example Ref.\cite{Lamoreaux}) gave rise to the study of the forces between objects of different geometries, dynamical phenomena (such as the dynamical Casimir effect and quantum friction, which both involve particle creation by moving boundaries) and heat transfer at the nanoscale\cite{VolokitinPersson}. Several discussions have taken place due to the introduction of models for real materials, exposing the discrepancies between the theoretical models and its constrast with experiments, such as the Drude vs Plasma controversy\cite{KlimchiMohMoste}. In fact, some of these discussions are currently open yet and are awaiting for solution.

Beyond all these advances, addressing non-equilibrium scenarios from both theory and experiments was a significant debt for a long time. In recent years, it gained great attention due to its potential applications\cite{BimonteEmigKardarKruger}. One of the first and remarkable works in the area is given by Ref.\cite{Antezza}, which study the Casimir force between two slabs characterized by arbitrary frequency-dependent permittivity functions. However, the approach employed is more an extension to non-equilibrium scenarios of the Lifshitz's method than a fully quantization scheme, and then takes part of the mentioned FQED approaches. As far as we know, a full quantum approach based on QFT for these situations was given subsequently in Refs.\cite{CTPGauge,TuMaRuLo}. In these works, the Huttner-Barnett (HB) microscopic model (consisting in the 3+1 electromagnetic (EM) field interacting with material polarizable bodies described by polarization degrees of freedom coupled to thermal baths in each point of space\cite{HuttnerBarnett}) is fully quantized by the closed time path (CTP) or Schwinger-Keldysh functional formalism\cite{CalzettaHu}. Implementing the influence functional to treat the field as an open quantum system, the field correlation can be calculated exactly and the result of Ref.\cite{Antezza} is recovered. Nevertheless, although both results agree, from a conceptual point of view, they present very subtle differences.

In Ref.\cite{Antezza} the steady (non-equilibrium) state is taken as an assumption and introduced in the calculations through implicitly assuming time-invariance for the Heisenberg equations when giving its Fourier transform as the starting point. Considering that, in fact, the Heisenberg equations are always subjected to initial conditions, which are discarded here by the `steadiness' assumption, in this sense we can say that the scheme is a `steady quantization scheme'. Moreover, the material bodies are described by the macroscopic frequency-dependent (complex) permittivity function and the fluctuations of the polarization sources inside, relating both by a fluctuation-dissipation relation containing the correct quantum thermal properties. As the relation is established separately for each point of each body, different temperatures for each body are easily introduced and the `steadiness' of the non-equilibrium scenario is quite reasonable and physically consistent. Then, non-equilibrium forces between two half-spaces and between two slabs (finite width) are analyzed. In the first case, the force results to be a sum of the contributions of each half-space, each one characterized by the respective temperature. For the two slabs case, the force present analog contributions (coming from each slabs) but, since the bodies' configuration is surrounded by vacuum, the radiation (coming from distant sources) impinging on the external surfaces also contribute to the force experienced by each slab. Therefore, contributions from this radiation enter the force. For these terms, the same expressions deduced exclusively for dissipative materials are employed for the vacuum (dissipationless) external regions, taking advantage implicitly of the continuity of the dissipative result with the dissipationless one when the damping constant is taken to zero (as it is done also in Ref.\cite{Maghrebi}). Apart from the calculations, all the framework and picture constitutes a FQED scheme at a steady scenario rather than a purely quantum one.

On the other hand, in Refs.\cite{CTPGauge,TuMaRuLo}, the situation arises in another way. The CTP formalism implemented does not assume a steady situation. On the contrary, is built for study the full time evolution from given initial conditions at time $t_{0}$, being closely related to the full solution of the Heisenberg equations. In this context, the steady situation emerges as the long-time limit of the full time evolution, i.e., by taking $t_{0}\rightarrow-\infty$ in all the expressions. Regarding the materials, in these works, a microscopic quantum model (similar to the HB model) is introduced to describe the internal dynamics of the degrees of freedom and thermal baths at each point of the dielectric bodies. Before the initial time $t_{0}$, all the parts of the total system (field and materials) are not interacting. At time $t_{0}$, all the interactions begin and the system start to evolve. Thus, the macroscopic EM properties of the material result from tracing out the internal degrees of freedom and the thermal baths during the interaction with the EM field. In this way, the steady situation is deduced and the fluctuation-dissipation relation results naturally from the open quantum system framework. In Ref.\cite{CTPGauge}, a correspondence between this quantum model and a stochastic description is fully demonstrated, exposing the existing connection between FQED and a fully quantum theory. In principle, three contributions are shown to appear for every case, each one associated to each field's sources resulting from the interactions present. One is associated to the thermal baths, another one to the internal degrees of freedom and the last one to the initial conditions of the field. In Ref.\cite{TuMaRuLo}, the formalism is applied to the study of the non-equilibrium half-space problem. The steady situation is deduced and it is shown that the only contribution present in this case is the one associated to the baths. Moreover, it gives exactly the same result as the one obtained from the FQED approach at Ref.\cite{Antezza}. However, related to the results of Ref.\cite{CTPScalar}, it is suggested from the calculations that for the case of slabs (finite width), in addition to the baths, the initial conditions would contribute to the long-time regime. This would be entirely related to the presence of infinite-size dissipationless regions, which causes that the initial free field fluctuations survive the damping in the dissipative material bodies to conform the vacuum modified modes when reaching the steady situation. In other words, the modified modes would be the steady result of the dynamical adaptation process of the initial free field modes (without boundaries) to the appearance of the material boundaries, giving a non-vanishing initial conditions' contribution at the long-time regime. In this train of thought, for the case of half-spaces, there is no initial conditions' contribution since, on the contrary, dissipation overcomes free-field fluctuations. In fact, in the framework of Ref.\cite{Antezza}, the initial conditions' contribution would matched with the distant-sources' contribution, although in this approach the radiation is related to the quantum fluctuations of the field already adapted to the boundaries. Moreover, it would also be matched with the homogeneous solution commented in Ref.\cite{Bechler}, obtained from a `steady path integral quantization' scheme. On the other hand, in a `steady canonical quantization' scheme, this homogeneous solution is quantized in a specific Hilbert space with its own creation and annihilation operators for the vacuum modes. All in all, the results of Refs.\cite{CTPGauge,TuMaRuLo,CTPScalar} are suggesting, indeed, that these operators are defined at the initial time, i.e., they are the creation and annihilation operators of the free field although they are used to quantize the homogeneous contribution when the boundaries are present.

In this sense, one of the main results of the present work is to show that all the mentioned suggestions and correspondences are true for arbitrary material configurations, validating the physical mechanism described above. For these purposes, we developed the canonical quantization version of Ref.\cite{CTPScalar}, where the material model mentioned above is considered in interaction with a scalar field. Clearly, the approach presented here can be extended to the EM case but for simplicity in the calculations we considered a one dimensional scalar field. We will proceed to solve the full dynamics of the total system, regarding the field correlation, which allow us to study the expectation values of the energy momentum tensor and, consequently, the Casimir force. We obtain the three mentioned contributions and we deduce its steady expressions by taking the long-time limit $t_{0}\rightarrow-\infty$. At this point, all the results are given for a general configuration. In this way, we are able to study the scalar field dynamics at any non-equilibrium scenario for this composite system. As far as we know, this kind of approach to non-equilibrium situations is completely new and its advantage is to be very suitable for computing the force and every quantity that can be obtained from the field operator. This is the main contribution of this work from a purely conceptual point of view.

Nevertheless, one of the practical advantages of this approach is that it gives the chance to study the quantities for non-classical initial states of the field. In quantum optics and cavity QED, squeezed states are the most common non-classical field states considered\cite{Dodonov1,Dodonov2}. For example, in quantum communications using this state, quantum ﬂuctuations of one of its quadrature components can be lower than that of the corresponding part in a coherent state, while the other component is higher. This gives the possibility of tuning the signal-to-noise ratios of its quadrature components, giving the chance to use one of the quadratures to absorb the quantum noise, while the other can be implemented to transmit an extremely well-deﬁned signal. In several applications, the quantum EM field inside a cavity is in a squeezed state. The squeezing can be generated through interacting with atoms inside the cavity\cite{ShBZheng}, which also open the way to study other different quantum phenomena as decoherence and teleportation\cite{VillaBoas,Guzman,JinHanZZYang,CaiKuang}. Hence, understanding the physics of the interactions pertaining the field inside a cavity and involving non-classical states is necessary to the development of new quantum technologies. Although the cavity is made of a good (but not perfect) conductor, it is common in the analysis to consider it as made of perfect materials, disregarding any corrections due to dissipation and non-equilibrium thermal effects\cite{BelgiornoLiberatiVisserSciama}. It is clear that the effectiveness of this approximation depends on the experimental conditions achieved. Therefore, it is of great interest to include these effects accurately. In this work we give light on how to address these calculations. We calculate the Casimir force between two dissipative material slabs (playing the role of cavity) when the field is in a squeezed state and each slabs has its own temperature. This greatly improve the results found in a previous work\cite{ZhengZheng}, where the Casimir force between two perfect conductor plates is calculated when the field is in a squeezed state. Moreover, our present result could be also implemented to enter the discussion given in Ref.\cite{Weigert} about the spatial properties of the squeezing of vacuum, including realistic details as thermal imbalance between the plates and the effect of dissipation, which escapes the scope of this work.

With the aim of focusing the main text of the work and the calculations on the mentioned results, we have left large calculations and deductions to several appendices at the end of the paper. The paper is organized as follows: in the next Section we describe the model and write the field equation with its solution. The full derivation of the field equation is left to Appendix \ref{AppA}. In Section \ref{CFO}, we study the long-time limit of the contributions to the field operator, with the main features of each one and the relation to other works. The analytical technique (including a general discussion of different configurations) along with the derivation of the long-time limit for each contribution are left to Appendices \ref{AppB}, \ref{AppC} and \ref{AppD}. Then, in Section \ref{CEMT}, it is shown that two contributions take part on the expectation values of the energy-momentum tensor at the steady state and each contribution is worked out. For the baths' contribution, thermal states are considered. For the initial conditions' contribution, the expectation values are computed both for thermal and also squeezed initial state of the field with the implementation of time-average. Section \ref{CFFWPC} is devoted to the calculation of the total Casimir force for the finite width plates configuration, deriving general expressions for each contribution the force. In Appendix \ref{AppE} we give the homogeneous solutions and the Green function needed for the calculations of the Section, while in Appendix \ref{AppF} we show how our result recovers the dissipationless limit and the Lifshitz formula at thermal equilibrium. In Section \ref{CBCFTSS}, we present a comparison between the Casimir forces obtained by taking a thermal state for the field and an squeezed one, when the baths keep the same temperature. Finally, Section \ref{FR} summarize our findings.

Throughout the paper, for simplicity, we have set $\hbar = k_{B} = c = 1$.

\section{Lagrangian Density and Field Equation}\label{LDFE}

With the aim of including effects of dissipation and noise in the evaluation of the Casimir energy or force, we will use the theory of open quantum systems, having in mind the paradigmatic example of the quantum Brownian motion (QBM) \cite{BreuerPetruccione}.

The model is a simplified version of the HB model, consisting of a system composed of two parts: a  massless scalar field and dielectric material which, in turn, are described by their internal degrees of freedom (a set of harmonic oscillators). Both sub-systems conform a composite system which is coupled to a second  set of harmonic oscillators, that plays the role of an external environment or thermal bath. For simplicity we will work in $1+1$ dimensions. In our toy model the massless field represents the electromagnetic field, and the first set of harmonic oscillators directly coupled to the scalar field represents the polarizable volume elements of the material.

Considering the usual interaction term between the electromagnetic field and the ordinary polarizable matter, the coupling between the field and the volume elements of the material will be taken as a current-type one, where the field couples to the velocity of variation of the volume elements' degrees of freedom. The coupling constant for this interaction is the electric charge $e$. We will also assume that there is no direct coupling between the field and the thermal bath. The Lagrangian density is therefore given by:

\begin{eqnarray}
\cal{L}&=&\cal{L}_{\phi}+\cal{L}_{S}+\cal{L}_{\phi-S}+\cal{L}_{B}+\cal{L}_{S-B}\nonumber\\&=&\frac{1}{2}~
\partial_{\mu}\phi~\partial^{\mu}\phi+4\pi\eta\left(\frac{1}{2}~m~\dot{r}^{2}_{x}(t)-\frac{1}{2}~m\omega_{0}^{2}~r^{2}_{x}(t)\right)+4\pi\eta
e~\phi(x,t)~\dot{r}_{x}(t)\nonumber\\&+&4\pi\eta\sum_{n}\left(\frac{1}{2}~m_{n}~\dot{q}_{n,x}^{2}(t)-\frac{1}{2}~m_{n}\omega_{n}^{2}~q_{n,x}^{2}(t)\right)-4\pi\eta\sum_{n}\lambda_{n}~q_{n,x}(t)~r_{x}(t),
\label{LagrangianaTOTAL}
\end{eqnarray}

\noindent where we have stressed the fact that $r$ and $q_{n}$ have a dependence on position as a label identifying the point of space as which they are located but without being dynamical variable (as it happens for the scalar field). It is clear that each atom interacts with a thermal bath placed at the same position. We have denoted by $\eta$ the density of the degrees of freedom of the volume elements. The constants $\lambda_{n}$ are the coupling constants between the volume elements and the bath oscillators. It is implicitly understood that Eq.(\ref{LagrangianaTOTAL}) represents the Lagrangian density inside the material, while outside the Lagrangian is given by the free field one.

The quantization of the theory is straightforward. It should be noted that the full Hilbert space of the model $H$, where the quantization is performed, is not only the field Hilbert space $H_{\phi}$ (as is considered in others works where the field is the only relevant degree of freedom), but also includes the Hilbert spaces of the volume elements' degrees of freedom $H_{\rm A}$ and the bath oscillators $H_{\rm B}$, in such a way that $H=H_{\phi}\otimes H_{\rm A}\otimes H_{\rm B}$. We will assume, as frequently done in the context of QBM,  that for $t<t_{0}$ the three parts of the systems are uncorrelated and not interacting. Interactions are turned on at $t=t_{0}$. Therefore, the initial conditions for the operators $\widehat{\phi}$, $\widehat{r}$ must be given in terms of operators acting in each part of the Hilbert space. The interactions will make that initial operators to  become operators over the whole space $H$. The initial density matrix of the total system is of the form:

\begin{equation}
\widehat{\rho}(t_{0})=\widehat{\rho}_{\rm IC}(t_{0})\otimes\widehat{\rho}_{\rm A}(t_{0})\otimes\widehat{\rho}_{\rm B},
\label{EstadoInicial}
\end{equation}

\noindent so, in principle, each part can be in any state.

Once the model of the interaction between the field and the matter is properly described, the equations of motion can be obtained. Solving the respective equations for the bath oscillators and the volume elements, an equation of motion for the field can be deduced. In appendix \ref{AppA}, this work is done in the context of the open quantum system's framework, showing that the field equation is given by:

\begin{equation}
\square\widehat{\phi}+\frac{\partial^{2}}{\partial t^{2}}\left[\int_{t_{0}}^{t}d\tau\chi_{x}(t-\tau)\widehat{\phi}(x,\tau)\right]=4\pi\eta
eC(x)\left[\dot{G}_{1}(t-t_{0})\widehat{r}_{x}(t_{0})+\dot{G}_{2}(t-t_{0})\frac{\widehat{p}_{x}(t_{0})}{m} +\int_{t_{0}}^{t}d\tau~\dot{G}_{2}(t-\tau)\frac{\widehat{F}_{x}(\tau-t_{0})}{m}\right],
\label{EcMovCampoTOTALKnoll}
\end{equation}

\noindent where $\chi_{x}(t)=\omega_{\rm Pl}^{2}~G_{2,x}(t)~C(x)$ is the susceptibility function with $\omega_{\rm Pl}^{2}=\frac{4\pi\eta e^{2}}{m}$ the plasma frecuency. It is worth noting that we have included an spatial label denoting the straightforward generalization to inhomogeneous media, where each point of the material can have different properties. Beyond this dependence, the boundaries of the material bodies enters through the spatial material distribution function $C$, which is zero in free space points. The regions filled (and the contours) with real material are defined by this function. This is clearly essential for the determination of the field's boundary conditions.

This equation (like all Heisenberg equations) is clearly subjected to initial conditions, in this case, free field conditions:

\begin{equation}
\widehat{\phi}(x,t_{0})=\int dk\left[\frac{1}{\omega_{k}}\right]^{\frac{1}{2}}\left(\widehat{a}_{k}(t_{0})~e^{i(kx-\omega_{k}t_{0})}+\widehat{a}_{k}^{\dag}(t_{0})~e^{-i(kx-\omega_{k}t_{0})}\right),
\label{CondIniCampoTOTAL}
\end{equation}

\begin{equation}
\dot{\widehat{\phi}}(x,t_{0})=\int dk\left[\frac{1}{\omega_{k}}\right]^{\frac{1}{2}}i\omega_{k}\left(-\widehat{a}_{k}(t_{0})~e^{i(kx-\omega_{k}t_{0})}+\widehat{a}_{k}^{\dag}(t_{0})~e^{-i(kx-\omega_{k}t_{0})}\right),
\label{CondIniDerCampoTOTAL}
\end{equation}

\noindent where $\widehat{a}_{k}(t_{0})$ and $\widehat{a}_{k}^{\dag}(t_{0})$ are the annihilation and creation operators for the free field at the initial time, and $\omega_{k}=|k|$.

To solve the equation, the retarded Green function can be employed, in such a way that the associated equation for $t>0$ reads:

\begin{eqnarray}
\square\mathfrak{G}_{\rm Ret}+\frac{\partial^{2}}{\partial t^{2}}\left(\int_{0}^{t}\chi_{x}(t-\tau)\mathfrak{G}_{\rm Ret}(x,x',\tau)d\tau\right)=0,
\label{EcMovGreen}
\end{eqnarray}

\noindent subjected to the following initial conditions:

\begin{eqnarray}
\mathfrak{G}_{\rm Ret}(x,x',t=0)=0~~~~~,~~~~~\dot{\mathfrak{G}}_{\rm Ret}(x,x',t=0)=-\delta(x-x'),
\label{CondInicialesGreen}
\end{eqnarray}

\noindent in such a way that the field can be written as:

\begin{eqnarray}
\widehat{\phi}(x,t)=&-&\int dx'~\dot{\mathfrak{G}}_{\rm Ret}(x,x',t-t_{0})~\widehat{\phi}(x',t_{0})-\int dx'~\mathfrak{G}_{\rm Ret}(x,x',t-t_{0})~\frac{\partial\widehat{\phi}}{\partial t}(x',t_{0})\nonumber\\
&-&\int_{t_{0}}^{t}dt'\int dx'~\mathfrak{G}_{\rm Ret}(x,x',t-t')~4\pi\eta eC(x')\left(\dot{G}_{1}(t'-t_{0})~\widehat{r}_{x'}(t_{0})+\dot{G}_{2}(t'-t_{0})~\frac{\widehat{p}_{x'}(t_{0})}{m}\right)\nonumber\\
&-&\int_{t_{0}}^{t}dt'\int dx'~\mathfrak{G}_{\rm Ret}(x,x',t-t')~4\pi\eta eC(x')\int_{t_{0}}^{t'}~d\tau~\dot{G}_{2}(t'-\tau)~\frac{\widehat{F}_{x'}(\tau-t_{0})}{m},
\label{OperadorCampoTOTALGreen}
\end{eqnarray}

\noindent which is the general solution for the field operator of the full time-dependent problem from given initial conditions.

As it expected, the field operator presents three parts, each one consisting in a operator acting in one of the three Hilbert spaces of the total Hilbert space.

\section{Contributions to the Field Operator}\label{CFO}

As it was recently stressed, the last equation of the preceding section means that the field operator begins at the initial time $t_{0}$ as an operator on $H_{\phi}$. Nevertheless, the switching-on of the interactions causes the field operator to become an operator on the full Hilbert space $H$ during the time evolution:

\begin{equation}
\widehat{\phi}(x,t)=\widehat{\phi}_{\rm IC}(x,t)\otimes\mathbb{I}_{\rm A}\otimes\mathbb{I}_{\rm B}+\mathbb{I}_{\phi}\otimes\widehat{\phi}_{\rm A}(x,t)\otimes\mathbb{I}_{\rm B}+\mathbb{I}_{\phi}\otimes\mathbb{I}_{\rm A}\otimes\widehat{\phi}_{\rm B}(x,t).
\end{equation}

As we are interested in evaluating the Casimir force in non-equilibrium but steady situations, we have to investigate the long-time limit ($t_{0}\rightarrow-\infty$) of these three contributions. Although the field operator will remain as an operator in the full Hilbert space $H$, as we will see, the Casimir force not necessarily contains contributions associated to each part of $H$. This will depends on the internal dynamics of the material and the initial state $\widehat{\rho}(t_{0})$ but also strongly on the boundaries' configuration considered, as it was commented in Refs.\cite{CTPScalar,TuMaRuLo}.

\subsection{Long-Time Limit of Initial Conditions' Contribution}\label{LTLICC}

Let us consider the field operator's contribution associated to the initial conditions. Since the initial conditions are written in terms of the creation and annihilation operators of the free field through Eqs.(\ref{CondIniCampoTOTAL}) and (\ref{CondIniDerCampoTOTAL}), the contribution can also be written in terms of these operators. Hence, the contribution splits into $\widehat{\phi}_{\rm IC}(x,t)=\widehat{\phi}_{\rm IC}^{(+)}(x,t)+\widehat{\phi}_{\rm IC}^{(-)}(x,t)$, with $\widehat{\phi}_{\rm IC}^{(-)}(x,t)=\left(\widehat{\phi}_{\rm IC}^{(+)}(x,t)\right)^{\dag}$ due to the fact that the retarded Green function $\mathfrak{G}_{\rm Ret}$ is real and the hermiticity of the initial free-field operators. Therefore, $\widehat{\phi}_{\rm IC}^{(+)}(x,t)$ is associated to the free-field annihilation operator $\widehat{a}_{k}(t_{0})$ and $\widehat{\phi}_{\rm IC}^{(-)}(x,t)$ to the free-field creation operator $\widehat{a}_{k}^{\dag}(t_{0})$. Moreover, given Eqs.(\ref{EcMovGreen}) and (\ref{CondInicialesGreen}), the initial conditions problem can be solved by Laplace transform. Therefore, the equation of motion for the retarded Green function's Laplace transform $\widetilde{\mathfrak{G}}_{\rm Ret}(x,x',s)$ is given by:

\begin{eqnarray}
\frac{\partial^{2}\widetilde{\mathfrak{G}}_{\rm Ret}}{\partial x^{2}}-s^{2}n_{x}^{2}(s)~\widetilde{\mathfrak{G}}_{\rm Ret}(x,x',s)=\delta(x-x'),
\label{EcMovGreenLAPLACE}
\end{eqnarray}

\noindent with $n_{x}(s)$ the refraction index at the point $x$:

\begin{equation}
n_{x}^{2}(s)\equiv1+\omega_{\rm Pl}^{2}~\widetilde{G}_{2,x}(s)~C(x).
\label{RefractionIndexX}
\end{equation}

The last equation is valid for every spatial dependence on the material properties and also for every configuration of boundaries. For a given material's distribution $C(x)$, the boundary conditions are determined through integrating the equation (It is worth noting that the Laplace transform $\widetilde{\mathfrak{G}}_{\rm Ret}$ turns out to be the Green function associated to the operator $\frac{\partial^{2}}{\partial x^{2}}-s^{2}n_{x}^{2}(s)$.).

Once we have obtained $\widetilde{\mathfrak{G}}_{\rm Ret}$ in the $s$-space, we can go back to the coordinate retarded Green function $\mathfrak{G}_{\rm Ret}$ via the Laplace anti-transform (or Mellin's formula, see Refs.\cite{KnollLeonhardt,SchiffLaplace}).



Therefore, the field operator can be written in terms of the Laplace transform of the retarded Green function:

\begin{equation}
\widehat{\phi}_{\rm IC}^{(+)}(x,t)=-\int dk\left(\frac{1}{\omega_{k}}\right)^{\frac{1}{2}}\widehat{a}_{k}(t_{0})~e^{-i\omega_{k}t_{0}}\int_{l-i\infty}^{l+i\infty}\frac{ds}{2\pi i}~e^{s(t-t_{0})}\left(s-i\omega_{k}\right)\int dx'~\widetilde{\mathfrak{G}}_{\rm Ret}(x,x',s)~e^{ikx'}.
\label{FieldOperatorGreenLaplaceAnnihilation}
\end{equation}

To find the Green function $\widetilde{\mathfrak{G}}_{\rm Ret}$, we use the technique found in Ref.\cite{Collin}, where the Green function of the Sturm-Liouville differential equation can be obtained from two solutions of the associated homogeneous equation, each one satisfying the boundary conditions on each side of the interval where the variable takes values. Then, we perform spatial integration for the general case of arbitrary number of interfaces. Finally, analyzing the complex-analytical properties given by the poles configuration, the long-time (steady) operator for this contribution can be worked out exactly by assuming causality as the only physical requirement. All this laborious task can be found in Appendix \ref{AppB}, where it is shown that the final and most general expression for the long-time operator is given by:

\begin{eqnarray}
\widehat{\phi}_{\rm IC}^{(+)}(x,t)\rightarrow\widehat{\phi}_{\rm IC}^{(+),\infty}(x,t)&=&\int dk\left[\frac{1}{\omega_{k}}\right]^{\frac{1}{2}}\widehat{a}_{k}(-\infty)\left[e^{-ikt}~\Theta(k)~\Phi_{-ik}^{>}(x)+e^{ikt}~\Theta(-k)~\left(\Phi_{-ik}^{<}(x)\right)^{*}\right]\nonumber\\
&&+~\left[\text{Time and spatial independent oscillatory term}\right].
\label{LongTimeFieldOperatorIC}
\end{eqnarray}

The first term of this expression is exactly the one suggested as an ansatz in the steady situation in Ref.\cite{Dorota1992}, based on the solution obtained for the dissipationless material case in Ref.\cite{Dorota1990}. This also includes and confirm the result shown in Ref.\cite{CTPScalar} for the initial condition contribution in the case of a single delta plate (which verifies the ansatz of Ref.\cite{Dorota1992}). It also includes the case of one thick plate analyzed in Ref.\cite{KnollLeonhardt}. In fact, this demonstration proves the general case based on canonical quantization scheme, extending to all the variety of situations. The functions $\Phi_{-ik}^{>}$ correspond to the modified modes for positive frequencies while $\left(\Phi_{-ik}^{<}\right)^{*}$ are the modes for the negative ones. The dynamical appearance and physics of these modes from a transient stage to a steady situation were commented briefly in Ref.\cite{CTPScalar} and deeper in Ref.\cite{TuMaRuLo}, although in the $3+1$ electromagnetic Lifshitz problem (two parallel half-spaces separated by a distance) analyzed in that work there was no modified modes at all. Here we have the same result for the Lifshitz problem of a $1+1$ scalar field. Clearly, the physics are the same. The modified modes appears in situations where there is an infinite-size dissipationless region, since the free fluctuation of the quantum field prevails over the dissipation in the finite regions occupied by materials, achieving a non-vanishing steady contribution at the long-time limit. Inversely, if there is no infinite-size dissipationless region, the initial conditions' contribution vanishes at the long-time limit, since dissipation overcomes free fluctuation in finite regions and the contribution is damped.

The second term, which is oscillatory and also time (and spatial) independent, will have no relevance on the calculation of the energy-momentum tensor expectation values since it involves time and spatial derivatives of the field operator.

As a final comment for this section, it is worth noting that the present demonstration is valid for every material represented by a refraction index $n$, which enters the Laplace transform of the retarded Green function's via Eq.(\ref{EcMovGreenLAPLACE}). Then, the deduction only stands on the Green function's properties, but without any restriction on the material model in addition to causality and physical consistence (which implies poles with non-positive real part). The information about the material is indeed contained in the specific form of the refraction index, which is the result of the interaction between the materials and the quantum field. Here, as it happens in Ref.\cite{KnollLeonhardt}, this is related to the definition of the susceptibility function $\chi$, obtained from solving the Heisenberg equations of motion for the material's degrees of freedom (volume elements plus thermal baths). In the CTP-integral formulation of Ref.\cite{CTPScalar} (and Ref.\cite{CTPGauge} for the electromagnetic version), the refraction index is directly related to the dissipation kernel generated by the material. However, as we have seen, that the initial conditions' contribution does not vanish in the steady regime is more related with the existence of infinite-size dissipationless regions rather than the material properties.

\subsection{Long-Time Limit of Volume Elements' Contribution}\label{LTLVEC}

Now, we consider the field operator's contribution associated to the volume elements. Since this contribution contains the volume elements' initial conditions $\{\widehat{r}_{x}(t_{0}),\widehat{p}_{x}(t_{0})\}$, the operator of this part always acts on the volume elements' Hilbert spaces. These operator can always be expressed in terms of the annihilation and creation operators:

\begin{equation}
\widehat{r}_{x}(t_{0})=\frac{1}{\sqrt{2m\omega_{0}}}\left(\widehat{b}_{0,x}^{\dag}(t_{0})+\widehat{b}_{0,x}(t_{0})\right),~~~~~\widehat{p}_{x}(t_{0})=i\sqrt{\frac{m\omega_{0}}{2}}\left(\widehat{b}_{0,x}^{\dag}(t_{0})-\widehat{b}_{0,x}(t_{0})\right).
\label{OperadorRt0enAyAdag}
\end{equation}

We can write this contribution as splitted in terms of the annihilation and creation operators of each volume element, $\widehat{\phi}_{\rm A}(x,t)=\widehat{\phi}_{\rm A}^{(+)}(x,t)+\widehat{\phi}_{\rm A}^{(-)}(x,t)$. From Eq.(\ref{OperadorCampoTOTALGreen}), we have:

\begin{equation}
\widehat{\phi}_{\rm A}^{(+)}(x,t)=-\int dx'~\frac{4\pi\eta eC(x')}{\sqrt{2m\omega_{0}}}~\widehat{b}_{0,x'}(t_{0})\int_{t_{0}}^{t}dt'~\mathfrak{G}_{\rm Ret}(x,x',t-t')\left(\dot{G}_{1}(t'-t_{0})-i\omega_{0}~\dot{G}_{2}(t'-t_{0})\right),
\label{PhiAFull}
\end{equation}

\noindent where, for simplicity, we have omitted the spatial labels on the volume elements' properties, such as the parameters (frequency, mass and density), or Green functions of the material ($G_{1,2}$). However, all the calculations will be valid considering this spatial dependence.

Again, the crucial point is to deduce the long-time limit for this operator contribution. The Green function can be written in terms of its Laplace transform and analyzing the poles configuration. The steady expression can be obtained by only involving assumptions related to causality. This work is realized at Appendix \ref{AppC}. Therefore, the long-time limit ($t_{0}\rightarrow-\infty$) of the operator is:

\begin{eqnarray}
\widehat{\phi}_{\rm A}^{(+)}(x,t)\longrightarrow\widehat{\phi}_{\rm A}^{(+),\infty}&=&-\frac{1}{2}\int dx'~\frac{4\pi\eta eC(x')}{\sqrt{2m\omega_{0}}}~\widehat{b}_{0,x'}(-\infty),
\label{LongTimeFieldOperatorA}
\end{eqnarray}

\noindent which is a time- and space-independent operator.

At first glance, is clear that the presence of the material distribution $C(x')$ causes that the integration is over the regions containing material.

However, since the energy-momentum tensor is constructed from expectation values of binary products of the field operator derivatives, the time and space independence of the long-time limit of the field operator makes the volume elements to not contribute at the steady situation for the physical dynamical quantities of interest. This is clearly in accordance with the results obtained in Refs.\cite{CTPScalar,TuMaRuLo,CTPGauge} for different situations. This also shows how the approximation considered in Refs.\cite{Dorota1992,LombiMazziRL}, where this contribution is discarded, is not correct at the operator level but turns out to be valid when calculating energy-momentum tensor expectation values.

As a final comment, it should be noted that the behaviour of the volume elements' contribution and the fact that it does not join the steady situation, is directly related with the material model considered. The dissipative dynamics of each volume element, considered effectively as a quantum brownian particle, makes the contribution vanishes at the long-time limit. Nevertheless, if the material model for the volume elements were taken to be quantum systems with non-dissipative dynamics, therefore a contribution at the steady situation would be present. This will be the case of the contribution of the baths of the next section. Moreover, this also be the case if, for example, we set the dissipation parameter $\gamma_{0}$ equal to zero, i.e., if we put every coupling constant between the baths and the volume elements equal to zero. The permittivity will be plasma-like and, in addition to the pole at $s=0$, two more poles with zero real part ($s=\pm i\omega_{0}$) will contribute to the volume elements' field operator that reach the steady situation. For the Drude model, the situation goes back to the case where $s=0$ is the only pole, although the expression of the long-time limit of the field operator changes slightly.

\subsection{Long-Time Limit of Thermal Baths' Contribution}\label{LTLTBC}

From Eq.(\ref{OperadorCampoTOTALGreen}), the baths' contribution is given by:

\begin{equation}
\widehat{\phi}_{\rm B}(x,t)=-\int_{t_{0}}^{t}dt'\int dx'~\mathfrak{G}_{\rm Ret}(x,x',t-t')~4\pi\eta eC(x')\int_{t_{0}}^{t'}~d\tau~\dot{G}_{2}(t'-\tau)~\frac{\widehat{F}_{x'}(\tau-t_{0})}{m}.
\label{PhiBFull}
\end{equation}

In contrast with how we proceeded for the other contributions, the full expression can be worked out now, instead of considering the annihilation contribution separated. However, the same methodological approach as for the other two contributions can be implemented. The crucial point is that, due to the dissipationless dynamics of the harmonic oscillators of the baths, the long-time limit of the solution is an operator that depends on both time and space. This work is shown in detail on Appendix \ref{AppD}, giving that the long-time contribution in this case reads:

\begin{equation}
\widehat{\phi}_{\rm B}(x,t)\longrightarrow\widehat{\phi}_{\rm B}^{\infty}(x,t)=\int dx'~\frac{4\pi\eta eC(x')}{m}\int_{-\infty}^{+\infty}\frac{d\omega}{2\pi}~e^{-i\omega t}~i\omega~\overline{G}_{2}(\omega)~\overline{\mathfrak{G}}_{\rm Ret}(x,x',\omega)~\widehat{\overline{F}}^{\infty}_{x'}(\omega),
\label{LongTimeFieldOperatorB}
\end{equation}

\noindent which has exactly the same form as the expression achieved in Ref.\cite{Antezza} for the EM field using a stochastic electrodynamics framework for the (quantum) Lifshitz problem.

It is also clear that the presence of the matter distribution $C$ makes the spatial integration to be carried out over the regions occupied by the material bodies, which in fact can be inhomogeneous since the local properties of the material can change in each point of space. This is why the integration over $x'$ contains every factor in the r.h.s., to eventually consider inhomogeneous materials. This last expression also has the time and space dependence suggested in Refs.\cite{Dorota1992,LombiMazziRL} for the so-called Langevin contribution associated to the baths and consisting in outgoing waves from the material bodies. This last feature is verified through the time dependence, while the space dependence is correctly supported by the spatial dependence of the field's retarded Green function transform $\overline{\mathfrak{G}}_{\rm Ret}$.

As a final comment to this section, it is clear that the transform of the stochastic force operator at the long-time limit $\widehat{\overline{F}}^{\infty}_{x'}(\omega)$ contains a limit on $t_{0}$ which seems to be oscillatory, however this will not enter the correlation expectation values of this operator, which are governed by the QBM theory.

\section{Contributions to the Energy-Momentum Tensor}\label{CEMT}

Finally, we have determined the long-time expressions for each part of the field operator, given in Eqs.(\ref{LongTimeFieldOperatorIC}), (\ref{LongTimeFieldOperatorA}) and (\ref{LongTimeFieldOperatorB}). Therefore, for $t_{0}\rightarrow-\infty$, the field operator reads:

\begin{equation}
\widehat{\phi}(x,t)\longrightarrow\widehat{\phi}^{\infty}(x,t)=\widehat{\phi}_{\rm IC}^{\infty}(x,t)\otimes\mathbb{I}_{\rm A}\otimes\mathbb{I}_{\rm B}+\mathbb{I}_{\phi}\otimes\widehat{\phi}_{\rm A}^{\infty}\otimes\mathbb{I}_{\rm B}+\mathbb{I}_{\phi}\otimes\mathbb{I}_{\rm A}\otimes\widehat{\phi}_{\rm B}^{\infty}(x,t),
\end{equation}

\noindent where we have stressed the fact that the volume elements' field operator at the steady situation does not depend on the spatial or temporal coordinates.

It is worth noting that a similar separation is considered, without rigorous demonstration, on Refs.\cite{Dorota1992,LombiMazziRL} about the field operator in the steady situation. In both works, the field operator does not contain any contribution from the volume elements. Based on the dissipative dynamics of the volume elements and the relaxation of its degrees of freedom, it is assumed that in the steady situation $\widehat{\phi}_{\rm A}^{\infty}\equiv0$. Here we show that this is not true and what it happens is that $\widehat{\phi}_{\rm A}^{\infty}$ is in fact independent of the spacetime coordinates. However, this has no direct implications on the calculation of the energy momentum tensor expectation values and forces, as we shall see, so the results of Refs.\cite{Dorota1992,LombiMazziRL} are correct. As we certainly have $\partial_{\mu}\widehat{\phi}_{\rm A}^{\infty}\equiv 0$, this implies that $\partial_{\mu}\widehat{\phi}^{\infty}=\partial_{\mu}\widehat{\phi}_{\rm IC}^{\infty}\otimes\mathbb{I}_{\rm A}\otimes\mathbb{I}_{\rm B}+\mathbb{I}_{\phi}\otimes\mathbb{I}_{\rm A}\otimes\partial_{\mu}\widehat{\phi}_{\rm B}^{\infty}$. It is worth noting that the derivative of the initial conditions' contribution on the last expression vanishes the time- and space-independent terms contained on $\widehat{\phi}_{\rm IC}^{\infty}$ (see Eq.(\ref{LongTimeFieldOperatorIC})).

Then, the expectation value of the components of the energy-momentum tensor operator at the steady situation will not contain any contribution of the volume elements independently of its initial state, what is in agreement with the results obtained by putting $\widehat{\phi}_{\rm A}^{\infty}\equiv0$ from the very beginning.



In the quantum theory, the expectation values of the energy-momentum tensor involves the correlation of the derivatives, which are the expectation values of symmetrized products. Therefore, for the components of the energy-momentum tensor operator we have:

\begin{equation}
\widehat{T}_{\mu\nu}(x_{1}^{\sigma},t_{0})\equiv\left(\delta_{\mu}^{~\gamma}\delta_{\nu}^{~\alpha}-\frac{1}{2}~\eta_{\mu\nu}\eta^{\gamma\alpha}\right)\frac{1}{2}\left(\partial_{\gamma}\widehat{\phi}(x_{1}^{\sigma})~\partial_{\alpha}\widehat{\phi}(x_{1}^{\sigma})+\partial_{\alpha}\widehat{\phi}(x_{1}^{\sigma})~\partial_{\gamma}\widehat{\phi}(x_{1}^{\sigma})\right),
\end{equation}

\noindent where we have stressed the dependence of this quantities on the initial time $t_{0}$.

Therefore, the long time limit of the components are straightforward:

\begin{eqnarray}
\widehat{T}_{\mu\nu}(x_{1}^{\sigma},t_{0})\longrightarrow\widehat{T}_{\mu\nu}^{\infty}(x_{1}^{\sigma})&=&\widehat{T}_{\mu\nu}^{{\rm IC},\infty}(x_{1}^{\sigma})\otimes\mathbb{I}_{\rm A}\otimes\mathbb{I}_{\rm B}+\mathbb{I}_{\phi}\otimes\mathbb{I}_{\rm A}\otimes\widehat{T}_{\mu\nu}^{{\rm B},\infty}(x_{1}^{\sigma})\\
&+&\left(\delta_{\mu}^{~\gamma}\delta_{\nu}^{~\alpha}-\frac{1}{2}~\eta_{\mu\nu}\eta^{\gamma\alpha}\right)\frac{1}{2}\Big[\partial_{\gamma}\widehat{\phi}_{\rm IC}^{\infty}(x_{1}^{\sigma})\otimes\mathbb{I}_{\rm A}\otimes\partial_{\alpha}\widehat{\phi}_{\rm B}^{\infty}(x_{1}^{\sigma})+\partial_{\alpha}\widehat{\phi}_{\rm IC}^{\infty}(x_{1}^{\sigma})\otimes\mathbb{I}_{\rm A}\otimes\partial_{\gamma}\widehat{\phi}_{\rm B}^{\infty}(x_{1}^{\sigma})\nonumber\\
&+&\partial_{\alpha}\widehat{\phi}_{\rm IC}^{\infty}(x_{1}^{\sigma})\otimes\mathbb{I}_{\rm A}\otimes\partial_{\gamma}\widehat{\phi}_{\rm B}^{\infty}(x_{1}^{\sigma})+\partial_{\gamma}\widehat{\phi}_{\rm IC}^{\infty}(x_{1}^{\sigma})\otimes\mathbb{I}_{\rm A}\otimes\partial_{\alpha}\widehat{\phi}_{\rm B}^{\infty}(x_{1}^{\sigma})\Big].\nonumber
\end{eqnarray}

As the derivatives of the baths' contribution to the field operator are obviously linear in the annihilation and creation operators of the baths and we are considering a thermal initial density matrix for the baths, we have that its expectation values are zero, i.e., $\langle\partial_{\gamma}\widehat{\phi}_{\rm B}^{\infty}(x_{1}^{\sigma})\rangle_{\rm B}\equiv {\rm Tr}_{\rm B}(\widehat{\rho}_{\rm B}\otimes\widehat{T}_{\mu\nu}^{{\rm B},\infty}(x_{1}^{\sigma}))=0$. This makes zero the expectation values of the second line of the last equation, independently on the field's initial state considered. Hence, for the expectation values of the components of the energy-momentum tensor reads:

\begin{equation}
\Big\langle\widehat{T}_{\mu\nu}^{\infty}(x_{1}^{\sigma})\Big\rangle=\Big\langle\widehat{T}_{\mu\nu}^{{\rm IC},\infty}(x_{1}^{\sigma})\Big\rangle_{\phi}+\Big\langle\widehat{T}_{\mu\nu}^{{\rm B},\infty}(x_{1}^{\sigma})\Big\rangle_{\rm B},
\label{ExpectationValueEnergyMomentumTensor}
\end{equation}

\noindent where $\Big\langle...\Big\rangle$ on the l.h.s. is the quantum expectation value over the total Hilbert space $H$, while $\Big\langle...\Big\rangle_{\phi}$ and $\Big\langle...\Big\rangle_{\rm B}$ on the r.h.s. are the quantum expectation values on the parts of the total Hilbert space associated to the field ($H_{\phi}$) and to the baths ($H_{\rm B}$) respectively.

Moreover, the last equation constitutes a generalization of the expression considered in Ref.\cite{LombiMazziRL} for the calculation of the pressure and also is in agreement with the separation of contributions deduced in Refs.\cite{CTPScalar,CTPGauge} for different specific situations studied through a functional integral approach. It is worth noting that in this case the thermal state of the baths ensure separation, regardless on the field's initial state. However, the same splitting can be achieved if the field have an initial thermal state, regardless on the state of the baths.

It is clear that the agreement between the calculations on Refs.\cite{Dorota1992,LombiMazziRL} (which mistakenly assume no contribution from the volume elements to the field operator) and the present ones relies on the fact that the physical quantities of interest are constructed from derivatives of the field operator. In this sense, if the field correlation 
could be measured directly, both approaches would differ due to the presence of the terms independent of the coordinates that would enter the field correlation, making the latter finally also depends on the volume elements' initial state.

\subsection{Thermal Baths' Contribution to the Energy-Momentum Tensor}\label{TBCEMT}

We start by calculating the contribution to the expectation values of the energy-momentum tensor associated to the thermal baths, given by the second term on Eq.(\ref{ExpectationValueEnergyMomentumTensor}):

\begin{equation}
\Big\langle\widehat{T}_{\mu\nu}^{{\rm B},\infty}(x_{1}^{\sigma})\Big\rangle_{\rm B}\equiv\left(\delta_{\mu}^{~\gamma}\delta_{\nu}^{~\alpha}-\frac{1}{2}~\eta_{\mu\nu}\eta^{\gamma\alpha}\right)\frac{1}{2}\Big\langle\left\{\partial_{\gamma}\widehat{\phi}_{\rm B}^{\infty}(x_{1}^{\sigma}),\partial_{\alpha}\widehat{\phi}_{\rm B}^{\infty}(x_{1}^{\sigma})\right\}\Big\rangle_{\rm B},
\label{ExpectationValueEnergyMomentumTensorBath}
\end{equation}

\noindent where $\{\widehat{A},\widehat{B}\}=\widehat{A}\widehat{B}+\widehat{B}\widehat{A}$ is the anticommutator of the operators $\widehat{A}$ and $\widehat{B}$.

Considering Eq.(\ref{LongTimeFieldOperatorB}), it derivative can be written as:

\begin{equation}
\partial_{\mu}\widehat{\phi}_{\rm B}^{\infty}(x,t)=\int dx'~\frac{4\pi\eta eC(x')}{m}\int_{-\infty}^{+\infty}\frac{d\omega'}{2\pi}~i\omega'\overline{G}_{2}(\omega')~e^{-i\omega't}\left[\delta_{\mu}^{~0}~(-i\omega')\overline{\mathfrak{G}}_{\rm Ret}(x,x',\omega')+\delta_{\mu}^{~1}~\partial_{x}\overline{\mathfrak{G}}_{\rm Ret}(x,x',\omega')\right]\widehat{\overline{F}}^{\infty}_{x'}(\omega').
\end{equation}




Therefore, the expectation value of the product of derivatives involves the correlation of the stochastic force operators.
These expectation values 
can be obtained from the definitions of the noise kernel and the stochastic force operator in Eqs.(\ref{QBMOperadorFdet}) and (\ref{QBMNucleoRuido}), as it is done in Ref.\cite{LombiMazziRL}:

\begin{equation}
\Big\langle\left\{\widehat{\overline{F}}^{\infty}_{x'}(\omega'),\widehat{\overline{F}}^{\infty}_{x''}(\omega'')\right\}\Big\rangle_{\rm B}=(2\pi)^{2}~\delta(x'-x'')~\frac{J(\omega')}{2\eta}~\coth\left(\frac{\beta_{\rm B}\omega'}{2}\right)\delta(\omega'+\omega'').
\end{equation}

Due to the delta functions, we obtain for the expectation value in the r.h.s. of Eq.(\ref{ExpectationValueEnergyMomentumTensorBath}):

\begin{eqnarray}
&&\Big\langle\left\{\partial_{\mu}\widehat{\phi}_{\rm B}^{\infty}(x,t),\partial_{\alpha}\widehat{\phi}_{\rm B}^{\infty}(x,t)\right\}\Big\rangle_{\rm B}=\int dx'~\omega_{\rm Pl}^{2}C(x')\int_{-\infty}^{+\infty}d\omega~\omega^{2}|\overline{G}_{2}(\omega)|^{2}2\pi\frac{J(\omega)}{m}\coth\left(\frac{\beta_{\rm B}\omega}{2}\right)\nonumber\\
&&\times\left[\delta_{\gamma}^{~0}~(-i\omega)\overline{\mathfrak{G}}_{\rm Ret}(x,x',\omega)+\delta_{\gamma}^{~1}~\partial_{x}\overline{\mathfrak{G}}_{\rm Ret}(x,x',\omega)\right]\left[\delta_{\alpha}^{~0}~i\omega~\overline{\mathfrak{G}}_{\rm Ret}^{*}(x,x',\omega)+\delta_{\alpha}^{~1}~\partial_{x}\overline{\mathfrak{G}}_{\rm Ret}^{*}(x,x',\omega)\right].
\end{eqnarray}

Considering Eq.(\ref{QBMNucleoAmortiguamiento}), it can be easily proved that $\omega\overline{\gamma}(\omega)=\frac{2\pi}{m}J(\omega)$. Moreover, from the definition of the refractive index below Eq.(\ref{EcMovGreenLAPLACE}) and considering a cut-off function without poles for the spectral density $J(\omega)$, it can be shown that:

\begin{equation}
\omega_{\rm Pl}^{2}~\omega\overline{\gamma}(\omega)|\overline{G}_{2}(\omega)|^{2}=2~{\rm Re}(n)~{\rm Im}(n),
\end{equation}

\noindent which is valid for every odd spectral density for any type of environment, in agreement with the results found in Ref.\cite{LombiMazziRL}.

Finally, for the expectation value of the components of the energy-momentum tensor, we have:

\begin{eqnarray}
\Big\langle\widehat{T}_{\mu\nu}^{{\rm B},\infty}(x)\Big\rangle_{\rm B}&=&\int dx'~C(x')\int_{-\infty}^{+\infty}d\omega~2\omega^{2}~{\rm Re}(n_{x'})~{\rm Im}(n_{x'})\coth\left(\frac{\beta_{{\rm B},x'}\omega}{2}\right)\nonumber\\
&\times&\Big(\left[\delta_{\mu}^{~0}~(-i\omega)\overline{\mathfrak{G}}_{\rm Ret}(x,x',\omega)+\delta_{\mu}^{~1}~\partial_{x}\overline{\mathfrak{G}}_{\rm Ret}(x,x',\omega)\right]\left[\delta_{\nu}^{~0}~i\omega~\overline{\mathfrak{G}}_{\rm Ret}^{*}(x,x',\omega)+\delta_{\nu}^{~1}~\partial_{x}\overline{\mathfrak{G}}_{\rm Ret}^{*}(x,x',\omega)\right]\nonumber\\
&-&\frac{1}{2}~\eta_{\mu\nu}\left[\omega^{2}|\overline{\mathfrak{G}}_{\rm Ret}(x,x',\omega)|^{2}-|\partial_{x}\overline{\mathfrak{G}}_{\rm Ret}(x,x',\omega)|^{2}\right]\Big),
\label{TMuNuB}
\end{eqnarray}

\noindent which does not depend on the time coordinate. On the other hand, there is still a spatial dependence in principle. Moreover, it should be noted that in the last expression we have included spatial labels for the material properties, denoting that the result is also valid for inhomogeneous materials.

This contribution of the baths to the energy-momentum tensor is in fact the $1+1$ scalar version and also the generalization (in terms of boundaries and inhomogeneity properties) of the expressions found in Refs.\cite{KnollLeonhardt,LombiMazziRL,CTPScalar,TuMaRuLo,Dorota1992,CTPGauge,Antezza,Maghrebi}, but this time deduced from a full canonical quantum procedure.

\subsection{Initial Conditions' Contribution to the Energy-Momentum Tensor}\label{ICCEMT}



We can now calculate the contribution to the energy-momentum tensor resulting from the initial conditions. With the aim of calculating the expectation values of the products of derivatives of the field operator, for simplicity we re-write Eq.(\ref{LongTimeFieldOperatorIC}) as:

\begin{eqnarray}
\widehat{\phi}_{\rm IC}^{\infty}(x,t)&=&\int dk\left[\frac{1}{\omega_{k}}\right]^{\frac{1}{2}}\left[\widehat{a}_{k}(-\infty)~e^{-i\omega_{k}t}~\Phi_{k}(x)+\widehat{a}_{k}^{\dag}(-\infty)~e^{i\omega_{k}t}\left(\Phi_{k}(x)\right)^{*}\right]\nonumber\\
&&+~\left[\text{Time and spatial independent oscillatory term}\right],
\end{eqnarray}

\noindent where we have to consider that $\Phi_{k}(x)=\Phi_{-ik}^{>}(x)$ for $k>0$ while $\Phi_{k}(x)=(\Phi_{-ik}^{<}(x))^{*}$ for $k<0$, and $\omega_{k}=|k|$.

This way, the derivative of the field operator is straightforward by considering that $\partial_{\mu}\left[e^{-i\omega_{k}t}\Phi_{k}(x)\right]=e^{-i\omega_{k}t}\left(\delta_{\mu}^{~0}(-i\omega_{k})~\Phi_{k}(x)+\delta_{\mu}^{~1}~\Phi_{k}^{'}(x)\right)$:

\begin{eqnarray}
\partial_{\mu}\widehat{\phi}_{\rm IC}^{\infty}(x^{\sigma})&=&\int dk\left[\frac{1}{\omega_{k}}\right]^{\frac{1}{2}}\Big[\widehat{a}_{k}(-\infty)~e^{-i\omega_{k}t}\left(\delta_{\mu}^{~0}(-i\omega_{k})~\Phi_{k}(x)+\delta_{\mu}^{~1}~\Phi_{k}^{'}(x)\right)\nonumber\\
&&+~\widehat{a}_{k}^{\dag}(-\infty)~e^{i\omega_{k}t}\left(\delta_{\mu}^{~0}~i\omega_{k}\left(\Phi_{k}(x)\right)^{*}+\delta_{\mu}^{~1}(\Phi_{k}^{'}(x))^{*}\right)\Big].
\end{eqnarray}



Considering that $\langle\{\widehat{a}_{k}^{\dag}(-\infty),\widehat{a}_{k'}^{\dag}(-\infty)\}\rangle_{\phi}=\langle\{\widehat{a}_{k}(-\infty),\widehat{a}_{k'}(-\infty)\}\rangle_{\phi}^{*}=2\langle\widehat{a}_{k}(-\infty)\widehat{a}_{k'}(-\infty)\rangle_{\phi}^{*}$ and also that $\langle\{\widehat{a}_{k}^{\dag}(-\infty),\widehat{a}_{k'}(-\infty)\}\rangle_{\phi}=2\langle\widehat{a}_{k}^{\dag}(-\infty)\widehat{a}_{k'}(-\infty)\rangle_{\phi}+\delta(k-k')=\langle\{\widehat{a}_{k}(-\infty),\widehat{a}_{k'}^{\dag}(-\infty)\}\rangle_{\phi}^{*}$, the expectation values of the components of the energy-momentum tensor are:



\begin{equation}
\Big\langle\widehat{T}_{\mu\nu}^{{\rm IC},\infty}(x_{1}^{\sigma})\Big\rangle_{\phi}=\Big\langle\widehat{T}_{\mu\nu}^{{\rm IC},\infty}(\mathbf{x}_{1})\Big\rangle_{\phi}^{\rm Vac}+\mathcal{T}_{\mu\nu}^{\rm State}(x_{1}^{\sigma}),
\label{TotalExpectationValueTMuNu}
\end{equation}

\noindent where $\Big\langle\widehat{T}_{\mu\nu}^{{\rm IC},\infty}(\mathbf{x}_{1})\Big\rangle_{\phi}^{\rm Vac}$ corresponds to the contribution associated entirely to vacuum fluctuations at zero temperature, which is always present, state- and (at least) time-independent and is given by:

\begin{equation}
\Big\langle\widehat{T}_{\mu\nu}^{{\rm IC},\infty}(\mathbf{x}_{1})\Big\rangle_{\phi}^{\rm Vac}=\int dk\frac{1}{\omega_{k}}{\rm Re}\Big[\left(\delta_{\mu}^{~0}(-i\omega_{k})\Phi_{k}+\delta_{\mu}^{~1}\Phi_{k}^{'}\right)\left(\delta_{\nu}^{~0}i\omega_{k})(\Phi_{k})^{*}+\delta_{\nu}^{~1}(\Phi_{k}^{'})^{*}\right)-\frac{\eta_{\mu\nu}}{2}\Big(\omega_{k}^{2}|\Phi_{k}|^{2}
-|\Phi_{k}^{'}|^{2}\Big)\Big],
\end{equation}

\noindent while $\mathcal{T}_{\mu\nu}^{\rm State}(x_{1}^{\sigma})$ corresponds to the specific contribution for the given initial state we consider for the field:

\begin{eqnarray}
\mathcal{T}_{\mu\nu}^{\rm State}(x_{1}^{\sigma})&=&\int dk\int dk'\left[\frac{1}{\omega_{k}\omega_{k'}}\right]^{\frac{1}{2}}2~{\rm Re}\Bigg[\Big\langle\widehat{a}_{k}(-\infty)~\widehat{a}_{k'}(-\infty)\Big\rangle_{\phi}~e^{-i(\omega_{k}+\omega_{k'})t}\Big[\Big(\delta_{\mu}^{~0}(-i\omega_{k})~\Phi_{k}+\delta_{\mu}^{~1}~\Phi_{k}^{'}\Big)\nonumber\\
&\times&\Big(\delta_{\nu}^{~0}(-i\omega_{k'})~\Phi_{k'}+\delta_{\nu}^{~1}~\Phi_{k'}^{'}\Big)+\frac{1}{2}~\eta_{\mu\nu}\Big(\omega_{k}\omega_{k'}~\Phi_{k}~\Phi_{k'}+\Phi_{k}^{'}~\Phi_{k'}^{'}\Big)\Big]\nonumber\\
&+&\Big\langle\widehat{a}_{k}^{\dag}(-\infty)~\widehat{a}_{k'}(-\infty)\Big\rangle_{\phi}~e^{i(\omega_{k}-\omega_{k'})t}\Big[\Big(\delta_{\mu}^{~0}i\omega_{k}\Phi_{k}+\delta_{\mu}^{~1}\Phi_{k}^{'}\Big)\left(\delta_{\nu}^{~0}(-i\omega_{k'})\left(\Phi_{k'}\right)^{*}+\delta_{\nu}^{~1}(\Phi_{k'}^{'})^{*}\right)\nonumber\\
&-&\frac{1}{2}~\eta_{\mu\nu}\Big(\omega_{k}\omega_{k'}\left(\Phi_{k}\right)^{*}\Phi_{k'}-(\Phi_{k}^{'})^{*}\Phi_{k'}^{'}\Big)\Big]\Bigg],
\label{TMuNuStatePart}
\end{eqnarray}

\noindent which in principle depends on all the spacetime coordinates.

All in all, depending on the initial state considered for the field, the different expectation values of the components of the energy-momentum tensor, which are expressed as a sum of a state-independent expectation value (corresponding to the vacuum fluctuations at zero temperature) and a state dependent term. For this last term, two important cases are to consider thermal and squeezed initial states.

\subsubsection{Thermal and Continuum-Single-Mode Squeezed States}\label{TCSMSS}

By considering a thermal initial state for the field, characterized by a temperature $\beta_{\phi}=1/T_{\phi}$, the expectation values of the products of annihilation and creation operators that appear in Eq.(\ref{TMuNuStatePart}) can be calculated straightforwardly:

\begin{equation}
\Big\langle\widehat{a}_{k}(-\infty)~\widehat{a}_{k'}(-\infty)\Big\rangle_{\phi}=0~~~~~,~~~~~\Big\langle\widehat{a}_{k}^{\dag}(-\infty)~\widehat{a}_{k'}(-\infty)\Big\rangle_{\phi}=N(\omega_{k})~\delta(k-k'),
\end{equation}

\noindent where $N(\omega_{k})=\frac{1}{e^{\beta_{\phi}\omega_{k}}-1}$ is the boson occupation number, i.e., the Bose-Einstein distribution.

In this case, $\mathcal{T}_{\mu\nu}^{\rm State}(x_{1}^{\sigma})$ simplifies giving the same integral as $\Big\langle\widehat{T}_{\mu\nu}^{{\rm IC},\infty}(\mathbf{x}_{1})\Big\rangle_{\phi}^{\rm Vac}$ but containing $2N(\omega_{k})$ in the integrand. In other words, $\mathcal{T}_{\mu\nu}^{\rm State}(x_{1}^{\sigma})$ turns out to be the thermal correction for the vacuum fluctuations at zero temperature, which is the expected result. Therefore, considering that $\coth\left(\frac{\beta_{\phi}\omega_{k}}{2}\right)=1+2N(\omega_{k})$, we have for Eq.(\ref{TotalExpectationValueTMuNu}):

\begin{eqnarray}
\Big\langle\widehat{T}_{\mu\nu}^{{\rm IC},\infty}(x_{1}^{\sigma})\Big\rangle_{\phi}&\equiv&\Big\langle\widehat{T}_{\mu\nu}^{{\rm IC},\infty}(\mathbf{x}_{1})\Big\rangle_{\phi}^{T_{\phi}}\nonumber\\
&=&\int dk\frac{1}{\omega_{k}}\coth\left(\frac{\beta_{\phi}\omega_{k}}{2}\right){\rm Re}\Big[\left(\delta_{\mu}^{~0}(-i\omega_{k})\Phi_{k}+\delta_{\mu}^{~1}\Phi_{k}^{'}\right)\left(\delta_{\nu}^{~0}i\omega_{k}\left(\Phi_{k}\right)^{*}+\delta_{\nu}^{~1}(\Phi_{k}^{'})^{*}\right)\nonumber\\
&&-\frac{\eta_{\mu\nu}}{2}\Big(\omega_{k}^{2}|\Phi_{k}|^{2}
-|\Phi_{k}^{'}|^{2}\Big)\Big],
\label{TMuNuICThermal}
\end{eqnarray}

\noindent which is, at least, a time-independent expression also. The dependence on the spatial coordinate is determined for each configuration through the introduction of the appropriate mode functions $\Phi_{k}$.


Considering a continuum-single-mode squeezed state for the field characterized by a squeezing parameter $\xi(k)=|\xi(k)|e^{i\varphi(k)}$ 
(see Refs.\cite{BarnettRadmore,BlowLoudon}):

\begin{equation}
|\xi(k)>=e^{-\frac{1}{2}\int_{-\infty}^{+\infty}dk\left(\xi(k)\widehat{a}_{k}^{\dag2}-\xi^{*}(k)\widehat{a}_{k}^{2}\right)}|0>
\label{ContinuumSingleModeSqueezedState}
\end{equation}

\noindent the expectation values of the products of annihilation and creation operators we have:

\begin{equation}
\Big\langle\widehat{a}_{k}(-\infty)~\widehat{a}_{k'}(-\infty)\Big\rangle_{\phi}=-~e^{i\varphi(k)}\sinh\left(|\xi(k)|\right)\cosh\left(|\xi(k)|\right)~\delta(k-k'),
\end{equation}

\begin{equation}
\Big\langle\widehat{a}_{k}^{\dag}(-\infty)~\widehat{a}_{k'}(-\infty)\Big\rangle_{\phi}=\sinh^{2}\left(|\xi(k)|\right)~\delta(k-k').
\end{equation}

In this case, the term of $\mathcal{T}_{\mu\nu}^{\rm State}(x_{1}^{\sigma})$ associated to $\langle\widehat{a}^{\dag}\widehat{a}\rangle$, combines with $\Big\langle\widehat{T}_{\mu\nu}^{{\rm IC},\infty}(\mathbf{x}_{1})\Big\rangle_{\phi}^{\rm Vac}$ because of the delta function $\delta(k-k')$. For the other term, $\delta(k-k')$ simplifies the expression, but results both spatial and time dependent.

Considering that $1+2\sinh^{2}\left(|\xi(k)|\right)=\cosh\left(2|\xi(k)|\right)$ and $2\sinh\left(|\xi(k)|\right)~\cosh\left(|\xi(k)|\right)=\sinh\left(2|\xi(k)|\right)$, we finally have:

\begin{equation}
\Big\langle\widehat{T}_{\mu\nu}^{{\rm IC},\infty}(x_{1}^{\sigma})\Big\rangle_{\phi}=\Big\langle\widehat{T}_{\mu\nu}^{{\rm IC},\infty}(\mathbf{x}_{1})\Big\rangle_{\phi}^{\rm Squeezed}+\mathcal{T}_{\mu\nu}^{\rm Squeezed}(x_{1}^{\sigma}),
\end{equation}

\noindent where the (at least) time-independent expectation value is given by:

\begin{eqnarray}
\Big\langle\widehat{T}_{\mu\nu}^{{\rm IC},\infty}(\mathbf{x}_{1})\Big\rangle_{\phi}^{\rm Squeezed}&=&\int dk\frac{1}{\omega_{k}}\cosh\left(2|\xi(k)|\right){\rm Re}\Big[\left(\delta_{\mu}^{~0}(-i\omega_{k})\Phi_{k}+\delta_{\mu}^{~1}\Phi_{k}^{'}\right)\left(\delta_{\nu}^{~0}i\omega_{k}\left(\Phi_{k}\right)^{*}+\delta_{\nu}^{~1}(\Phi_{k}^{'})^{*}\right)\nonumber\\
&&-\frac{\eta_{\mu\nu}}{2}\Big(\omega_{k}^{2}|\Phi_{k}|^{2}-|\Phi_{k}^{'}|^{2}\Big)\Big],
\end{eqnarray}

\noindent and the second term:

\begin{eqnarray}
\mathcal{T}_{\mu\nu}^{\rm Squeezed}(x_{1}^{\sigma})&=&-\int dk\frac{1}{\omega_{k}}\sinh\left(2|\xi(k)|\right){\rm Re}\Bigg[e^{i\varphi(k)}e^{-i2\omega_{k}t}\Big[\Big(\delta_{\mu}^{~0}(-i\omega_{k})~\Phi_{k}+\delta_{\mu}^{~1}~\Phi_{k}^{'}\Big)\Big(\delta_{\nu}^{~0}(-i\omega_{k})\Phi_{k}+\delta_{\nu}^{~1}\Phi_{k}^{'}\Big)\nonumber\\
&&+\frac{1}{2}\eta_{\mu\nu}\Big(\omega_{k}^{2}~\Phi_{k}^{2}+\Phi_{k}^{'2}\Big)\Big]\Bigg].
\end{eqnarray}

This last expression can be taken one step further through time-averaging. For this steady quantities defined at the long-time limit, the time-average for a quantity $A$ is given by:

\begin{equation}
\langle\langle A\rangle\rangle_{t}=\lim_{\tau\rightarrow+\infty}\frac{1}{\tau}\int_{-\tau/2}^{\tau/2}d\tau'~A(\tau').
\end{equation}

Therefore, for the time dependence in the integrand of $\mathcal{T}_{\mu\nu}^{\rm Squeezed}$, we have $\langle\langle e^{-i2\omega_{k}t}\rangle\rangle_{t}=\lim_{\tau\rightarrow+\infty}\frac{\sin(\omega_{k}\tau)}{\omega_{k}\tau}=0$ for $k\neq 0$ and $\langle\langle 1\rangle\rangle_{t}=1$ for $k=0$. However, for this last case, the rest of the integrand of $\mathcal{T}_{\mu\nu}^{\rm Squeezed}$ vanishes since $\Phi_{k}$ is given by sums of $e^{\pm ikx}$. Finally, we obtain:

\begin{equation}
\left\langle\left\langle\mathcal{T}_{\mu\nu}^{\rm Squeezed}\right\rangle\right\rangle_{t}=0.
\end{equation}

It should be noted that this result is due to the oscillatory time-dependence for every $k$. If the modes have another dependence, this last time average could be different from zero. However, for boundary conditions on the spatial coordinate only, the time dependence is in general oscillatory for a field in dielectric media and, therefore, the time average vanishes.

All in all, for both cases, after taking the time-average, the expectation value of the energy-momentum tensor can be written as:

\begin{eqnarray}
\left\langle\left\langle\Big\langle\widehat{T}_{\mu\nu}^{{\rm IC},\infty}(x_{1}^{\sigma})\Big\rangle_{\phi}\right\rangle\right\rangle_{t}&=&\int dk\frac{1}{\omega_{k}}~\mathfrak{F}(k)~{\rm Re}\Big[\left(\delta_{\mu}^{~0}(-i\omega_{k})\Phi_{k}+\delta_{\mu}^{~1}\Phi_{k}^{'}\right)\left(\delta_{\nu}^{~0}i\omega_{k}\left(\Phi_{k}\right)^{*}+\delta_{\nu}^{~1}(\Phi_{k}^{'})^{*}\right)\nonumber\\
&&-\frac{\eta_{\mu\nu}}{2}\Big(\omega_{k}^{2}|\Phi_{k}|^{2}-|\Phi_{k}^{'}|^{2}\Big)\Big],
\label{TMuNuFdeK}
\end{eqnarray}

\noindent having $\mathfrak{F}(k)=\coth\left(\frac{\beta_{\phi}\omega_{k}}{2}\right)$ for a thermal state and $\mathfrak{F}(k)=\cosh\left(2|\xi(k)|\right)$ for a squeezed state. It should be noted that other initial states could provide another time and spatial dependence for the expectation values of the components of the energy-momentum tensor, complicating the time-average procedure and opening a new type of formulae for the force. The case of continuum-single-mode squeezed states, given the expectation values for annihilation and creation operators, turns out to be very simple as we will see in the next section. Due to the time-average, the forces for thermal and continuum-single-mode squeezed states can be written as particular cases of the last expression. Similar results can be obtained for the case of continuum-two-mode squeezed states characterized by a frequency $\Omega$ (see Ref.\cite{BarnettRadmore} for the quantum state), which after time-averaging the expressions results in basically the same result. On the other hand, other initial states, as continuum-coherent states for example, would present more complicated expressions, including double integration over the frequency, since the expectation values for the products of annihilation and creation operators do not include a Dirac delta function $\delta(k-k')$.

\section{Casimir Force for Finite Width Plates Configuration}\label{CFFWPC}

Once we have given expressions for the expectation values of the components of the energy-momentum operator, we proceed to calculate the Casimir force between two homogeneous plates of finite width $d$ and different materials ($n_{L}$ and $n_{R}$ for the left and right plates respectively) separated by a distance $a$. With this aim, to calculate the force over one of the plates we substract the field's pressures in each side of the plate. In our case, the pressure is given by the expectation value of the $xx-$component of the energy-momentum tensor operator. Moreover, as this expectation value splits in two contributions, the same is true for the Casimir force (as it happens in Refs.\cite{BreuerPetruccione,KnollLeonhardt,LombiMazziRL,CTPScalar,TuMaRuLo,Dorota1992,CTPGauge,Maghrebi}). Therefore, we have:

\begin{equation}
F_{\rm C}=\langle\widehat{T}_{xx}\rangle^{\rm Ext}-\langle\widehat{T}_{xx}\rangle^{\rm Int}=\langle\widehat{T}_{xx}^{\rm IC}\rangle^{\rm Ext}+\langle\widehat{T}_{xx}^{\rm B}\rangle^{\rm Ext}-\langle\widehat{T}_{xx}^{\rm IC}\rangle^{\rm Int}-\langle\widehat{T}_{xx}^{\rm B}\rangle^{\rm Int}=F_{\rm C}^{\rm IC}+F_{\rm C}^{\rm B},
\label{ForceCasimir}
\end{equation}

\noindent where the superscript `$\rm Int$' denotes the region of the vacuum gap between the plates and `$\rm Ext$' denotes the region outside the plates configuration adjacent to the respective plate under consideration. It is clear that in the last expression, for the initial conditions' contribution to the energy-momentum tensor, we are considering the time-averaged expression given in Eq.(\ref{TMuNuFdeK}).

Considering Eqs.(\ref{TMuNuB}) and (\ref{TMuNuFdeK}), for the full calculation of each contribution, we need both the mode functions (or homogeneous solutions) $\Phi$ and the transform of the Green function $\mathfrak{G}_{\rm Ret}$ for the two plates configuration. These expressions are given in Appendix \ref{AppE}. Then, we can easily calculate the contribution to the Casimir force acting on the left plate when the field is in a thermal or continuum-single-mode squeezed state through Eqs.(\ref{TMuNuFdeK}) and (\ref{ForceCasimir}), obtaining:

\begin{equation}
F_{\rm C}^{\rm IC}[a,d,\mathfrak{F}]=\int_{0}^{+\infty}dk~k~\mathfrak{F}(k)\left[1+\left|R_{-ik}^{>}\right|^{2}+\left|T_{-ik}\right|^{2}-\left|C_{-ik}^{>}\right|^{2}-\left|D_{-ik}^{>}\right|^{2}-\left|C_{-ik}^{<}\right|^{2}-\left|D_{-ik}^{<}\right|^{2}\right],
\label{CasimirForceICFull}
\end{equation}

\noindent which is an extension of the results found in Refs.\cite{LombiMazziRL,Dorota1990,Dorota1992,Dorota1993}. The explicit dependence on $\mathfrak{F}$ denotes that the same expression is valid for both thermal and squeezed states. Nevertheless, while for the former, $\mathfrak{F}(k)$ is always an even function, for the latter, for an even $\mathfrak{F}(k)$ is required an even $\xi(k)$, which was assumed for obtaining the last result.

It should be noted that due to the dissipation present in this scenario, $\left|R_{-ik}^{>}\right|^{2}+\left|T_{-ik}\right|^{2}\neq 1$ and $|r_{i}|^{2}+|t_{i}|^{2}\neq 1$.

For the contribution of the baths, employing the last expressions and replacing them into Eq.(\ref{TMuNuB}) to obtain the contribution as Eq.(\ref{ForceCasimir}) states, it is straightforward to obtain:

\begin{eqnarray}
F_{\rm C}^{\rm B}\left[a,d,\beta_{\rm B,L},\beta_{\rm B,R}\right]&=&\int_{0}^{+\infty}d\omega~\frac{\omega}{8}\left|T_{-i\omega}\right|^{2}\Bigg[\coth\left[\frac{\beta_{\rm B,L}\omega}{2}\right]\frac{|n_{\rm L}+1|^{2}}{|n_{\rm L}|^{2}}\frac{1}{|t_{\rm R}|^{2}|t_{\rm L}|^{2}}\Bigg({\rm Re}(n_{\rm L})(1-e^{-2\omega{\rm Im}(n_{\rm L})d})\nonumber\\
&\times&\Big[|(1-r_{\rm L}r_{\rm R}e^{i2\omega a})(1-r_{n_{\rm L}}r_{\rm L})-r_{n_{\rm L}}r_{\rm R}t_{\rm L}^{2}e^{i2\omega a}|^{2}-|(1-r_{\rm L}r_{\rm R}e^{i2\omega a})(r_{\rm L}-r_{n_{\rm L}})+r_{\rm R}t_{\rm L}^{2}e^{i2\omega a}|^{2}\nonumber\\
&+&|t_{\rm L}|^{2}(1+|r_{\rm R}|^{2})(1-|r_{n_{\rm L}}|^{2})\Big]+2~{\rm Im}(n_{\rm L})~{\rm Im}\Big[\left(e^{i2\omega{\rm Re}(n_{\rm L})d}-1\right)\Big([(1-r_{\rm L}r_{\rm R}e^{i2\omega a})(1-r_{n_{\rm L}}r_{\rm L})\nonumber\\
&-&r_{n_{\rm L}}r_{\rm R}t_{\rm L}^{2}e^{i2\omega a}][(1-r_{\rm L}^{*}r_{\rm R}^{*}~e^{-i2\omega a})(r_{\rm L}^{*}-r_{n_{\rm L}}^{*})+r_{\rm R}^{*}t_{\rm L}^{*2}~e^{-i2\omega a}]+|t_{\rm L}|^{2}(1+|r_{\rm R}|^{2})r_{n_{\rm L}}\Big)\Big]\Bigg)\nonumber\\
&+&\coth\left[\frac{\beta_{\rm B,R}\omega}{2}\right]\frac{|n_{\rm R}+1|^{2}}{|n_{\rm R}|^{2}}\left[1-\frac{1+|r_{\rm L}|^{2}}{|t_{\rm L}|^{2}}\right]\Bigg({\rm Re}(n_{\rm R})(e^{2\omega{\rm Im}(n_{\rm R})d}-1)[1-|r_{n_{\rm R}}|^{2}]-2~{\rm Im}(n_{\rm R})\nonumber\\
&\times&{\rm Im}\left[r_{n_{\rm R}}(e^{i2\omega{\rm Re}(n_{\rm R})d}-1)\right]\Bigg)\Bigg].
\label{CasimirForceBathFull}
\end{eqnarray}

Therefore, considering the last expression and Eq.(\ref{CasimirForceICFull}), the non-equilibrium force experienced by the left plate of a Casimir configuration is given by:

\begin{equation}
F_{\rm C}\left[a,d,\mathfrak{F},\beta_{\rm B,L},\beta_{\rm B,R}\right]=F_{\rm C}^{\rm IC}\left[a,d,\mathfrak{F}\right]+F_{\rm C}^{\rm B}\left[a,d,\beta_{\rm B,L},\beta_{\rm B,R}\right].
\label{CasimirForceFull}
\end{equation}

All in all, this is the Casimir force for a non-equilibrium scenario consisting in two plates of finite width $d$ and different materials. It is, in fact, the generalization of the results found in Refs.\cite{LombiMazziRL,Dorota1992}. As it happens in those situations, it is expected that our new result is regularized by the dependence of the reflection coefficient on $k$, which ensures convergence by including a natural cut-off in the model considered.

Moreover, two important limit-cases can be recovered from this general expression. One is the force for the case of materials without dissipation (real frequency-independent refractive indexes) and the other one is the Lifshitz formula, which is the force between two half-spaces (infinite width) at thermal equilibrium. Appendix \ref{AppF} is devoted to show how these results can be recovered from our general non-equilibrium expressions.

\section{Comparison Between Casimir Forces for Thermal and Squeezed States}\label{CBCFTSS}

Once we have calculated the Casimir force between two finite width plates in Eqs.(\ref{CasimirForceICFull}), (\ref{CasimirForceBathFull}) and (\ref{CasimirForceFull}) for both situations, when the initial state for the field is thermal or squeezed, while the baths in each point of the plates are always characterized by its own temperature value. Therefore, the comparisons between both cases and between our result and previous ones are mandatory. Although we have general formulae valid even for the case of different temperatures in each slab, in the present work we focus on the comparison between different states of the field, while keeping the same temperature on both plates (and equal to the temperature of the field when considering the thermal state for it). For simplicity, analyzing full non-equilibrium scenarios, including different temperatures between the parts and squeezed states is left as pending future work.

In a previous work \cite{ZhengZheng}, the Casimir force between perfect conductor plates was calculated when the EM field is in a squeezed state. Given the perfect material, the plates enter as boundary conditions on the quantum field. As the material of the plates does not present any internal dynamics, the system of interest (the field) is not an open system for this case.

Therefore, the quantization of the system is based directly on quantizing the modes of the field, confined to the space between the plates. This implies that only a Hilbert space for the field is required for the corresponding quantum theory.

As the field is confined in the transverse direction to the plates, the transverse component of the wave vector of each mode is discretized. Then, this is inherited by the eigenfrequencies of the problem. As in our calculation, the squeezed state enters when computing the expectation values of products of the creation and annihilation operators. This results in the factor $\cosh(2|\xi_{m}|)$, equivalent to ours but logically discretized due to the allowed modes for this idealized case (being the subscript $m$ the label of the discrete modes here). This was done in Ref.\cite{ZhengZheng} where, for simplicity, one of the cases analyzed was to take a constant squeezing for all the modes, $\xi_{m}=\xi$ for every $m$. Therefore, the factor $\cosh(2|\xi_{m}|)$ is a constant and the force results to be the well-known Casimir's result times $\cosh(2|\xi|)$.

In our case, we are considering a one-dimensional scalar field instead of a full EM field. However, the respective limit of perfect conductors can be addressed from our general formulae. First step is to erase any internal dynamics in order to remove the dissipation from the result. As it is commented in the first section of Appendix \ref{AppF}, this is achieved by setting $\overline{\gamma}_{\rm L,R}(\omega)\equiv0$ (which gives $F_{\rm C}^{\rm B}\equiv0$) and taking the zeroth order of the permittivity functions. Therefore, the Casimir force is given by the initial conditions' contribution only through Eq.(\ref{CasimirForceNODiss}). Therefore, for the case of an initial squeezed state, we have to put $\mathfrak{F}(k)=\cosh(2|\xi(k)|)$. If we also impose a constant squeezing for all the modes, we obtain:

\begin{eqnarray}
F_{\rm C}\left[a,d,\xi=\text{const.}\right]\Big|_{\text{No Diss}}&=&\cosh(2|\xi|)\int_{0}^{+\infty}dk~k\left[2-\frac{[|t_{\rm L}|^{2}(1+|r_{\rm R}|^{2})+|t_{\rm R}|^{2}(1+|r_{\rm L}|^{2})]}{|1-r_{\rm L}r_{\rm R}~e^{i2\omega a}|^{2}}\right]\nonumber\\
&=&\cosh(2|\xi|)~F_{\rm C}^{T_{\phi}=0}\left[a,d\right]\Big|_{\text{No Diss}},
\end{eqnarray}

\noindent where $F_{\rm C}^{T_{\phi}=0}\left[a,d\right]\Big|_{\text{No Diss}}$ is the Casimir force between two slabs of materials without dissipation at zero temperature, which is in agreement with Ref.\cite{Dorota1993}.

This result is analog to the one obtained in Ref.\cite{ZhengZheng} but to the case of two slabs of materials without dissipation. Moreover, for the case of perfect-conductor plates, following the procedure given in Ref.\cite{Dorota1990} for the respective limit (analyzing the behaviour of the integrand), it can be easily shown that $F_{\rm C}^{T_{\phi}=0}\left[a,d\right]\Big|_{\text{No Diss}}$ can be reduced to the corresponding expression for the Casimir force between perfect conductors. Therefore, the one-dimensional scalar version of the result given in Ref.\cite{ZhengZheng} is fully achieved.

\begin{figure}
\centering
\includegraphics[width=1.0\linewidth]{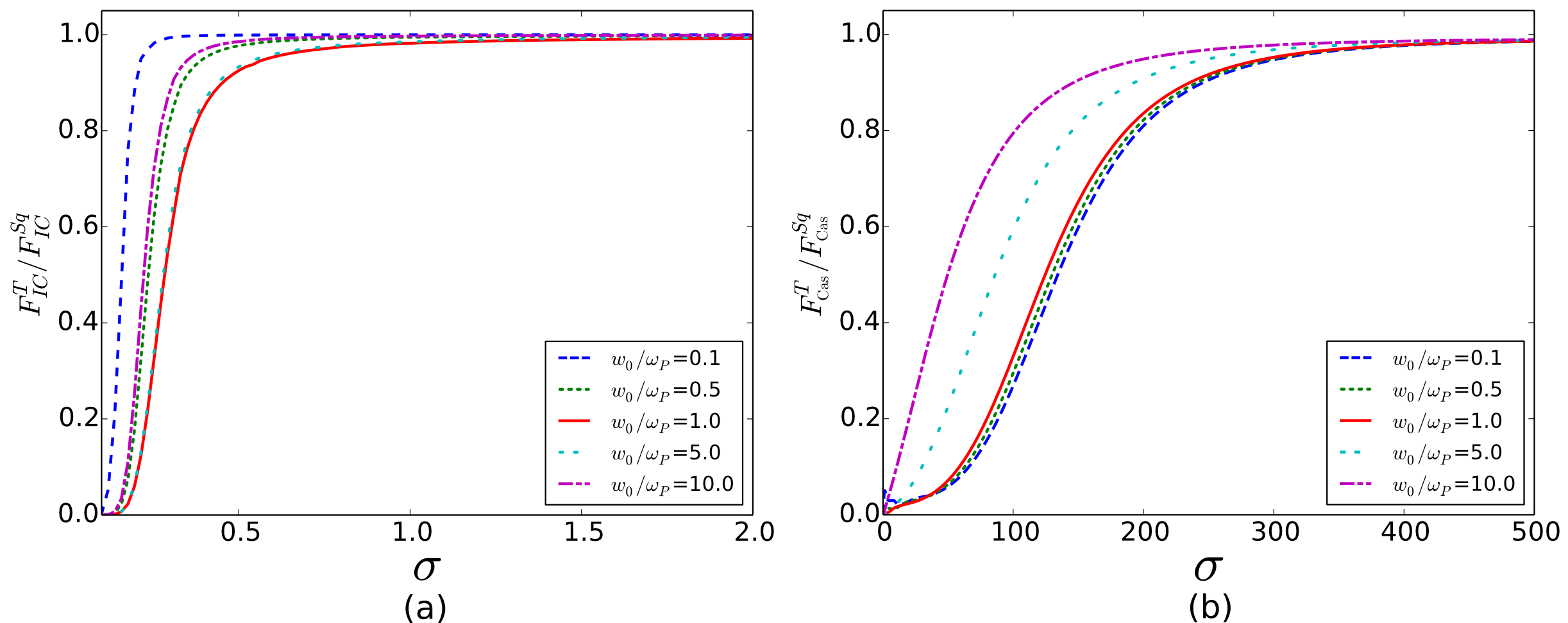}
\caption{a. Ratio of the initial conditions' contributions for a thermal state with $T_{\phi}=300K$ and a continuum-single-mode squeezed state as a function of the bandwidth $\sigma$ for different values of the center frequency $\Omega_{0}$ (compared with the plasma frequency $\omega_{\rm Pl}$). b. Ratio of the total forces for a thermal state with $T_{\phi}=300K$ and a continuum-single-mode squeezed state as a function of $\sigma$ for different $\Omega_{0}$. Parameters are $\gamma_{\rm L,R}=10^{-1}/a$, $\omega_{0,i}=10/a$, $\omega_{\rm Pl,i}=10/a$, $d/a=10^{2}$, $a=100\rm nm$, $\sigma$ is given in units of $a$.}
\label{fig:1}
\end{figure}

Nevertheless, considering constant squeezing for all the modes of the field is too idealized. On the one hand, in cavity QED, it is usual to consider that in perfect conductor cavities, only one of the discrete modes of the field is squeezed\cite{VillaBoas,Guzman,JinHanZZYang,CaiKuang}. On the other hand, in our case, the cavity is formed by dielectric slabs, so it is reasonable to expect that the squeezing is not limited to a unique mode. If we want to squeeze one mode of frequency $\Omega_{0}$, we expect to effectively squeeze this mode but also the modes around it contained in a bandwidth $\sigma$ centered in $\Omega_{0}$. We assume for simplicity that the squeezing parameter $\xi(k)$ for all the modes contained in the bandwidth (for $\Omega_{0}-\frac{\sigma}{2}<k<\Omega_{0}+\frac{\sigma}{2}$) is the same and given by $1/\sigma$ (there is no significant changes in the analysis if we choose for the squeezing $\textit{const.}/\sigma$, being $\textit{const.}$ any real number). This choice is because we want the squeezing parameter $\xi$, as a function of $k$, to recover a squeezed state only for the mode $\Omega_{0}$ when the bandwidth $\sigma$ approaches $0$. In other words, we are demanding that $\xi(k)\rightarrow\delta(k-\Omega_{0})$ when $\sigma\rightarrow0$, in such a way that Eq.(\ref{ContinuumSingleModeSqueezedState}) result a squeezed state only for the frequency $\Omega_{0}$. Accordingly, comparing the initial conditions' contributions for a squeezed state with this squeezing distribution $\xi_{\sigma,\Omega_{0}}(k)$ and for a thermal state with $T_{\phi}=300K$ (room temperature), with in both cases considering ohmic thermal baths, we obtain Fig.\ref{fig:1}a for different values of $\Omega_{0}$ as a function of $\sigma$ (see Ref.\cite{BreuerPetruccione}).

As it can be seen, the curves in Fig.\ref{fig:1}a are always below 1, implying that the contribution for the squeezed state is always greater than for the thermal state for any values of $\sigma$ and $\Omega_{0}$. This is expected because the presence of the factor $\cosh(2|\xi(k)|)$ in the integrand, combined with the chosen squeezing distribution, makes the integration unfolds, but all the frequencies keep contributing to the value of the integral. For the frequencies outside the bandwidth $\sigma$ centered at $\Omega_{0}$ we have that $\cosh(2|\xi(k)|)\equiv1$ since $\xi(k)\equiv0$, while for the frequencies belonging to the bandwidth, $\cosh(2|\xi(k)|)$ takes a constant value greater than 1. Therefore, the final value of the initial conditions' contribution for the squeezed state results greater than for the thermal state.

However, as the bandwidth becomes larger, the value of the squeezing given by $1/\sigma$ results smaller. Therefore, for large $\sigma$, the squeezing approaches to zero and the value of the contribution for the squeezed state gets closer to the thermal value. At first glance, this seems natural since the initial conditions contribution at zero temperature result from putting $\mathfrak{F}\equiv1$ for every $k$. But, in our case, we are comparing the contribution for a squeezed state with the value for a thermal one with $T_{\phi}=300K$ instead of zero temperature. However, the integrand of the initial conditions' contribution does not change significantly for different values of temperature because $k\left[1+\left|R_{-ik}^{>}\right|^{2}+\left|T_{-ik}\right|^{2}-\left|C_{-ik}^{>}\right|^{2}-\left|D_{-ik}^{>}\right|^{2}-\left|C_{-ik}^{<}\right|^{2}-\left|D_{-ik}^{<}\right|^{2}\right]$ in Eq.(\ref{CasimirForceICFull}) is different from zero when the thermal factor $\coth\left(\frac{\beta_{\phi}k}{2}\right)$ is close to 1. This means that the initial conditions' contribution is insensitive to the chosen temperature. Thus, the ratio between the initial conditions' contributions for a thermal state with $T_{\phi}=300K$ and for a squeezed state with large bandwidth $\sigma$ (which approaches the result with $T_{\phi}=0K$) is logically close to 1.

A similar argument also explains the fact that the curves in Fig.\ref{fig:1}a are not ordered according to its values of $\Omega_{0}$ for the values of $\sigma$ shown in the figure. The point is that for a fixed $\sigma$, different values of $\Omega_{0}$ enhance (through the squeezing factor) different parts of the spectrum of the integrand. Therefore, the final value of the contribution to the force strongly depends on the chosen $\Omega_{0}$ for a narrow bandwidth $\sigma$. However, for large $\sigma$, the enhanced parts of the integrand are very similar, independently of the chosen value of $\Omega_{0}$ and the final value of the contributions approach and get ordered according to it. This can be seen at the end of the curves and it was checked for higher values of $\sigma$.

On the other hand, the behaviour of the curves for very small $\sigma$ is explained by another feature of the squeezing distribution $\xi(k)$. As the bandwidth becomes narrower, the contribution for the modes inside it gets greatly enhanced since the squeezing parameter is given by $1/\sigma$. Therefore, although the number of enhanced modes is lower, its contributions to the integral grow strongly. Therefore, the final contributions to the force also grow, getting numerically divergent values when the bandwidth approaches zero. Therefore, the ratio for the contributions of the force in the different situations tends to zero. However, it should be noted that this is in fact a numerical limitation instead of a correct result. The limit of $\sigma\rightarrow0$ is well-defined analytically. As we mentioned before, the choice of the squeezing distribution was to describe an imperfect squeezing of a mode of frequency $\Omega_{0}$ characterized by a bandwidth $\sigma$, in such a way that the perfect squeezing on the mode $\Omega_{0}$ only could be obtained through the limit of null bandwidth. If we like to obtain this result, we should go to Eq.(\ref{ContinuumSingleModeSqueezedState}) and set $\xi(k)=\delta(k-\Omega_{0})$. Then, the full calculation of the initial conditions' contribution is straightforward, obtaining that it corresponds to the evaluation of the integrand at Eq.(\ref{CasimirForceICFull}) on the chosen frequency $\Omega_{0}$, which a well-defined finite result. It is clear that this value strongly depends on the chosen frequency, however it remains always finite.

Beyond all these features, although the final value of the initial conditions' contribution to the force seems to vary significantly, it does for a small interval of $\sigma$ (until $0,75$ in unit of $a$ for the given curves). Then, it seems to be insensitive to the changes of $\sigma$ for a wide range of values. This is because we are comparing the initial conditions' contributions only. However, while the baths' contributions to the total force (based on the expelled field by the materials of the plates) tends to separate the plates, the initial conditions' contribution tends to attract them. Then, the total force results from a substraction which is more sensitive to the changes of $\sigma$. This is observed in Fig.\ref{fig:1}b, where the approach to an asymptotic value (also from below) of the ratio of the total forces for the squeezed and thermal states occurs for values of $\sigma$ four orders of magnitude larger (until $350$ in units of $a$ approximately).

It should be noted that for the baths in each plate, the temperatures were all equal to $300K$. Thus, for the thermal case, the total force is the one at thermal equilibrium.

Contrary to what we have when comparing the initial conditions' contributions separately, the curves are ordered by its value of $\Omega_{0}$. For the large-$\sigma$ behaviour, the asymptotic value is close to 1 but it is not equal to it. This shows that the ratio of the total forces is sensitive to the chosen temperature. On the other hand, the behaviour of the curves for small $\sigma$ present the same troubles that mentioned for the other figure. Again, the limit $\sigma\rightarrow0$ can be studied analytically.

All in all, the ratio between the total forces when the initial conditions' contributions are different seems to allow tuning the force by modifying the squeezing properties (the bandwidth $\sigma$ and the frequency $\Omega_{0}$) of the field state in a wide range of values. As the ratio is smaller than 1, the tuning could intensify the (always attractive) total force from a minimum value defined by the temperature of the plates.

\section{Final Remarks}\label{FR}

In this work we have study several aspects of the Casimir forces in different out of equilibrium situations, analyzing its value in some remarkable situations, elucidating the steady dynamics of different non-trivial scenarios for the chosen model.

Specifically, from a conceptual point of view, we have achieved a well-defined and consistent approach for the study of the Casimir forces at completely general scenarios. We developed a first-principles canonical quantization formalism for the study of the interaction between a quantum scalar field and dissipative-material polarizable bodies. These are locally modeled as complex quantum systems in each point of space, formed by a quantum harmonic oscillators describing the volume elements of the polarizable body which are coupled to its own thermal bathm, as it commonly used since Ref.\cite{HuttnerBarnett}. In analogy to the results obtained through the CTP integral formulation given in Refs.\cite{CTPGauge,CTPScalar}, we have obtained a full equation of motion for the field interacting effectively with the material bodies. However, as in Ref.\cite{KnollLeonhardt}, the equation stands for the quantum field operator and (as every Heisenberg equation) is subjected to initial conditions, which for simplicity we took as free field operators. The solution for this equation is given in terms of the Green function and all the sources included in the problem that generate field due to the interactions between the different parts of the total system. It is important to remark that the interactions are switched-on at a given initial time $t_{0}$, being a sudden start analog to a quench. For arbitrary time, we obtain three contributions to the field operator, each one associated to a specific part of the total system. One is associated to the volume elements, another one is associated to the baths and, finally, there is a contribution entirely related to the field initial conditions. As the parts of the total system are initially free, the field operator separate in three contributions, each one acting on the respective Hilbert space ($H_{\phi}\otimes H_{\rm A}\otimes H_{\rm B}$). This splitting is critical for study the long-time limit ($t_{0}\rightarrow-\infty$) of each contribution separately.

We achieved expressions for the long-time contribution to the field operator of each part of the total system, showing that at the steady situation only two of them contribute to the expectation values of the energy-momentum tensor components. There is no long-time contribution from the volume elements. Moreover, we showed that the baths' contribution to the long-time field operator matches with the one considered in Ref.\cite{Antezza} from a FQED approach, while the one associated to the initial conditions have the form of the homogeneous solutions considered in previous works for other steady quantization schemes. In fact, this contribution also agree with the result obtained in Ref.\cite{CTPScalar} and the mechanism described in Ref.\cite{TuMaRuLo}, where this was suggested but not fully demonstrated. Here, we describe clearly how the existence of infinite-size regions without dissipation for the field are directly related to a non-vanishing long-time contribution of the initial conditions. All in all, as Refs.\cite{CTPScalar,CTPGauge} but this time from a canonical quantization scheme, our general result close a conceptual hole between the full quantum theory and the FQED approaches. The matching between both and how FQED is related to quantum fluctuations is clearly explained.

Once exploring the general features of the developed approach, the other main result of the work is based on taking advantage of the results obtained for the expectation values of the energy-momentum tensor components of general material configurations. We proceed to fully calculate the force between two slabs of different material and temperatures but the same width. The initial conditions' contribution was evaluated for two types of states, thermal and continuum-single-squeezed. For the latter, time-average was required to determine the value of the force and write the two cases in a unified way. Then, for the squeezed case, we showed how to recover the results at Ref.\cite{ZhengZheng} of the force between perfect conductor plates. However, our result is the out of equilibrium generalized version to dissipative materials and squeezed states for the field. Therefore, we proceed to compare this case with the result at thermal equilibrium, taking $T_{\phi}=300K$. From Fig.\ref{fig:1}a, we showed that the initial conditions' contribution is sensitive to the change on the bandwidth of squeezed modes $\sigma$ only for small values of the bandwidth, regardless on the centering frequency $\Omega_{0}$. Moreover, the contribution results insensitive when changing the temperature. Nevertheless, in Fig.\ref{fig:1}b, we showed that when comparing the total force for each case, the sensitivity is increased in four orders of magnitude when changing the values of $\sigma$. We also showed that the total force is also sensitive when changing the temperature. All in all, considering a constant squeezing for given $\sigma$ and $\Omega_{0}$, we showed that the change of $\sigma$ allow to tune the value of the (always attractive) Casimir force from a minimum value defined by the temperature of the plates. This could be of great interest for application in technological improvements.

As a final comment, it should be noted that this results can be easily extended to the three-dimensional case, and also to the EM field, where two kind of modes enter, the evanescent and the propagating. For our case of a one-dimensional scalar field, we only deal with propagating modes. However, this is left as pending work. Moreover, out of thermal equilibrium forces and heat transfer can be studied from the present results. Also the situation of changing suddenly the distance between the plates, which is related to the dynamical Casimir effect. This can be achieved by employing the long-time field operator for a given configuration of the plates as the initial condition for a new configuration. This is also left as pending future work.

\acknowledgments

We would like to thank Fernando C. Lombardo for useful comments and discussions about the manuscript, and to Pablo M. Poggi and Vincenzo Gianinni for their invaluable help in the preparation of the numerical code employed for the figures and the discussion on the analysis. This work is supported by CONICET, Argentina.

\appendix
\section{Deducing the Field Equation}\label{AppA}

In this section, we show how Eq.(\ref{EcMovCampoTOTALKnoll}) is rigorously obtained given the chosen model. Starting from the Lagrangian (\ref{LagrangianaTOTAL}), it is easy to derive the Heisenberg equations of motion for the different operators. They are are given by:

\begin{equation}
\widehat{p}_{n,x}=m_{n}~\dot{\widehat{q}}_{n,x},~~~~~\dot{\widehat{p}}_{n,x}=-m_{n}\omega_{n}^{2}~\widehat{q}_{n,x}+\lambda_{n}~\widehat{r}_{x},
\label{EcMovTOTALA}
\end{equation}

\begin{equation}
\widehat{p}_{x}=m~\dot{\widehat{r}}_{x}+e~\widehat{\phi},~~~~~\dot{\widehat{p}}_{x}=-m\omega_{0}^{2}~\widehat{r}_{x}+\sum_{n}\lambda_{n}~\widehat{q}_{n,x},
\label{EcMovTOTALC}
\end{equation}

\begin{equation}
\square\widehat{\phi}=4\pi\eta e~\dot{\widehat{r}}_{x},
\label{EcMovTOTALE}
\end{equation}

\noindent where the operators $\widehat{p}_{x}$ and $\widehat{p}_{n,x}$ are the conjugate momentum operators associated to the operators $\widehat{r}_{x}$ and $\widehat{q}_{n,x}$ respectively.

Combining the first two equation, we get:

\begin{equation}
m_{n}~\ddot{\widehat{q}}_{n,x}+m_{n}\omega_{n}^{2}~\widehat{q}_{n,x}-\lambda_{n}~\widehat{r}_{x}=0.
\label{EcMovQN}
\end{equation}

As usual in the context of QBM,  we solve the equations for the operators $\widehat{q}_{n,x}$ taking $\widehat{r}_{x}$ as a source, and replace the solutions into the pair of Eq.(\ref{EcMovTOTALC}). In this way, the microscopic degrees of freedom in the material satisfy a Langevin-like equation of the form:

\begin{equation}
\dot{\widehat{p}}_{x}=-m\omega_{0}^{2}~\widehat{r}_{x}-m~\frac{d}{dt}\left(\int_{t_{0}}^{t}d\tau~\gamma(t-\tau)~\widehat{r}_{x}(\tau)\right)+\widehat{F}_{x}(t-t_{0}),
\label{EcMovRYP}
\end{equation}

\noindent where the damping kernel $\gamma$ and the stochastic force operator $\widehat{F}_{x}$ are the same as the ones of QBM (see \cite{BreuerPetruccione} for a general and complete view). They are given by:

\begin{equation}
\gamma(t)=\frac{2}{m}\int_{0}^{+\infty}d\omega~\frac{J(\omega)}{\omega}~\cos(\omega t),
\label{QBMNucleoAmortiguamiento}
\end{equation}

\begin{equation}
\widehat{F}_{x}(t-t_{0})=\sum_{n}\frac{\lambda_{n}}{\sqrt{2m_{n}\omega_{n}}}\left(e^{-i\omega_{n}(t-t_{0})}~\widehat{b}_{n,x}(t_{0})+e^{i\omega_{n}(t-t_{0})}~\widehat{b}_{n,x}^{\dag}(t_{0})\right).
\label{QBMOperadorFdet}
\end{equation}

Here $\widehat{b}_{n,x}(t_{0})$ and $\widehat{b}_{n,x}^{\dag}(t_{0})$ are the annihilation and creation operators associated to $\widehat{q}_{n,x}(t_{0})$, and $J(\omega)$ is the spectral density that characterizes the environment, which gives the number of oscillators in each frecuency for given values of the coupling constants $\lambda_{n}$ (see Refs.\cite{BreuerPetruccione,LombiMazziRL} for more details).





At equilibrium, the stochastic force operator in Eq.(\ref{QBMOperadorFdet}) and the damping kernel $\gamma$ in Eq.(\ref{QBMNucleoAmortiguamiento}) are not independent. The statistical properties of the stochastic force operator are given by the dissipation and noise kernels:

\begin{equation}
D(t-t')\equiv
i\left\langle\left[\widehat{F}(t);\widehat{F}(t')\right]\right\rangle=i\left[\widehat{F}(t);\widehat{F}(t')\right]=-2\int_{0}^{+\infty}d\omega~J(\omega)~\sin\left(\omega(t-t')\right),
\label{QBMNucleoDisipacion}
\end{equation}

\begin{equation}
D_{1}(t-t')\equiv
\left\langle\left\{\widehat{F}(t);\widehat{F}(t')\right\}\right\rangle=2\int_{0}^{+\infty}d\omega~ J(\omega)~\coth\left(\frac{\beta_{\rm B}\omega}{2}\right)\cos\left(\omega(t-t')\right),
\label{QBMNucleoRuido}
\end{equation}

\noindent which are the formal quantum open systems generalization of the relations employed in Ref. \cite{KnollLeonhardt} for general environments and arbitrary temperature. Note that only the noise kernel $D_{1}$ involves the environmental temperature $\beta_{\rm B}=1/T_{\rm B}$ as a parameter. Considering Eq.(\ref{QBMNucleoAmortiguamiento}) and Eq.(\ref{QBMNucleoDisipacion}), it is easy to show that:

\begin{equation}
\frac{d}{dt}\Big(\gamma(t-s)\Big)=-\frac{1}{m}~D(t-s),
\label{QBMNucleoAmortiguamientoProp1}
\end{equation}

\noindent which relates the damping kernel $\gamma$ to the statistical properties of the stochastic force operator $\widehat{F}$.

All in all, the set of equations to solve now are Eqs.(\ref{EcMovTOTALC}), (\ref{EcMovTOTALE}), and (\ref{EcMovRYP}).

It is possible to obtain a formal solution for the operators $\widehat{r}_{x}(t)$ by  considering the field $\phi$ as a source for the equation. This solution generalizes the crude approximation made in Ref. \cite{KnollLeonhardt} for the evolution of the microscopic degrees of freedom in the mirrors. It is given by:

\begin{equation}
\widehat{r}_{x}(t)=G_{1}(t-t_{0})~\widehat{r}_{x}(t_{0})+G_{2}(t-t_{0})~\dot{\widehat{r}}_{x}(t_{0})+\frac{1}{m}\int_{t_{0}}^{t}d\tau~G_{2}(t-\tau)\left(\widehat{F}_{x}(\tau-t_{0})-e~\dot{\widehat{\phi}}(x,\tau)\right),
\label{SolucionQBMAcoplado}
\end{equation}

\noindent where $G_{1,2}$ are the Green functions associated to the QBM equation that satisfy:

\begin{align}
G_{1}(0)=1\;\;\;\;\text{,}\;\;\;\;\dot{G}_{1}(0)=0,
\label{CondInicialesG1}
\end{align}

\begin{align}
G_{2}(0)=0\;\;\;\;\text{,}\;\;\;\;\dot{G}_{2}(0)=1,
\label{CondInicialesG11}
\end{align}

\noindent for which, the Laplace transforms are given by

\begin{equation}
\widetilde{G}_{n}(z)=\frac{z^{2-n}}{z^2+\omega_{0}^2+z~\widetilde{\gamma}(z)},
\label{TransformadaG1}
\end{equation}

\noindent with $n=1,2$ and where $\widetilde{\gamma}$ is the Laplace transform of the damping kernel. Note that,  given these conditions, one can  prove that $G_{1}(t)=\dot{G}_{2}(t)$.

Inserting this solution into Eq.(\ref{EcMovTOTALE}), we obtain the following equation for the field operator given in Ref.\cite{LombiMazziRL}:

\begin{equation}
\square\widehat{\phi}+\frac{4\pi\eta
e^2}{m}\int_{t_{0}}^{t}d\tau~G_{1}(t-\tau)\dot{\widehat{\phi}}(x,\tau)=4\pi\eta
e\left[\dot{G}_{1}(t-t_{0})\widehat{r}_{x}(t_{0})+G_{1}(t-t_{0})\dot{\widehat{r}}_{x}(t_{0}) +\frac{1}{m}\int_{t_{0}}^{t}d\tau~G_{1}(t-\tau)\widehat{F}_{x}(\tau-t_{0})\right],
\label{EcMovCampoTOTAL}
\end{equation}

\noindent which can also be rewritten as in Ref.\cite{KnollLeonhardt}, finally arriving to Eq.(\ref{EcMovCampoTOTALKnoll}).

\section{Steady Initial Conditions' Contribution from Complex Analysis}\label{AppB}

This section is devoted to show how Eq.(\ref{LongTimeFieldOperatorIC}) is obtained for the long-time operator of the initial conditions' contribution.

We begin by considering Eq.(\ref{FieldOperatorGreenLaplaceAnnihilation}) as starting point. Following Ref.\cite{Collin}, we can write the desired Green function as:

\begin{equation}
\widetilde{\mathfrak{G}}_{\rm Ret}(x,x',s)=\frac{\Phi^{<}_{s}(x_{<})~\Phi^{>}_{s}(x_{>})}{W_{s}(x')},
\label{GreenFunctionCollin}
\end{equation}

\noindent where $x_{>}$ ($x_{<}$) is the bigger (smaller) value between $x$ and $x'$, $\Phi^{<}_{s}$ ($\Phi^{>}_{s}$) is the homogeneous solution associated to Eq.(\ref{EcMovGreenLAPLACE}) satisfying only the boundary condition on the left (right) limit of the variable value's interval and $W_{s}(x)=\Phi^{<}_{s}(x)~\frac{d\Phi^{>}_{s}}{dx}-\frac{d\Phi^{<}_{s}}{dx}~\Phi^{>}_{s}(x)$ is the wronskian of the solutions (which has to be independent of $x$).

At this point, we go a step further and give a demonstration for the long-time limit of this operator for a completely general case. We will consider the case of $N-1$ interfaces separating different materials characterized by refractive index $n_{\alpha}$, with $\alpha=1,2,...,N-1,N$ for a given positive integer $N$. However, beyond the regions imposed by the $N-1$ interfaces, the construction of the Green function $\widetilde{\mathfrak{G}}_{\rm Ret}(x,x',s)$ always contains a jolt associated to the relation between the field and source points ($x,x'$ respectively). Given that the field point $x$ is chosen in the $m-$region, the spatial integral over the source point $x'$ in Eq.(\ref{FieldOperatorGreenLaplaceAnnihilation}) separates in $N+1$ integrals, $N-1$ integrals associated to each region not containing $x$, and two more associated to the splitting of the $m-$region due to the jolt in the relation between $x$ and $x'$. Therefore:

\begin{equation}
\int dx'\longrightarrow\left[\int_{-\infty}^{x_{1}}+\int_{x_{1}}^{x_{2}}+...+\int_{x_{m-2}}^{x_{m-1}}+\int_{x_{m-1}}^{x}+\int_{x}^{x_{m}}+\int_{x_{m}}^{x_{m+1}}+...+\int_{x_{N-1}}^{x_{N}}+\int_{x_{N}}^{+\infty}\right]dx',
\label{IntSplit}
\end{equation}

\noindent where $\{x_{i}\}$ are the positions of the $N$ interfaces.

Then, considering the construction of the Green function through Eq.(\ref{GreenFunctionCollin}), the first $m$ integrals in Eq.(\ref{IntSplit}) are of the form:

\begin{equation}
\int_{x_{\alpha-1}}^{x_{\alpha}}dx'~\widetilde{\mathfrak{G}}_{\rm Ret}(x,x',s)~e^{ikx'}=\frac{\Phi_{s}^{>}(x)}{W_{s}}\int_{x_{\alpha-1}}^{x_{\alpha}}dx'~\Phi_{s}^{<}(x')~e^{ikx'},
\end{equation}

\noindent for every $\alpha=1,...,m$ by considering $x_{0}=-\infty$ for the lower limit in the integral with $\alpha=1$ and $x_{m}=x$ in the upper limit of the last integral ($\alpha=m$).

The last $N-m+2$ integrals in Eq.(\ref{IntSplit}) are of the form:

\begin{equation}
\int_{x_{\alpha-1}}^{x_{\alpha}}dx'~\widetilde{\mathfrak{G}}_{\rm Ret}(x,x',s)~e^{ikx'}=\frac{\Phi_{s}^{<}(x)}{W_{s}}\int_{x_{\alpha-1}}^{x_{\alpha}}dx'~\Phi_{s}^{>}(x')~e^{ikx'},
\end{equation}

\noindent with $\alpha=m,...,N+1$ by considering $x_{m-1}=x$ for the first integral and $x_{N+1}=+\infty$ for the upper limit of the last integral.

By considering that the homogeneous solutions in each region can be, in general (except for the first and last region as we will see below), written as the superposition of waves travelling to the right and to the left, i.e., $\Phi_{s}^{\lessgtr}(x)=\sum_{j=+,-}K_{s}^{\lessgtr}(j,\alpha)~e^{jsn_{\alpha}(s)x}$, where $K_{s}^{\lessgtr}(j,\alpha)$ are the coefficients resulting from the appropriate boundary conditions. Therefore, we have:

\begin{eqnarray}
\int dx'~\widetilde{\mathfrak{G}}_{\rm Ret}(x,x',s)~e^{ikx'}&=&\frac{\Phi_{s}^{>}(x)}{W_{s}}\sum_{j=+,-}j\Bigg[\sum_{\alpha=1}^{m-1}\frac{K_{s}^{<}(j,\alpha)}{\left(sn_{\alpha}(s)+j~ik\right)}\left(e^{(jsn_{\alpha}(s)+ik)x_{\alpha}}-e^{(jsn_{\alpha}(s)+ik)x_{\alpha-1}}\right)\nonumber\\
&+&\frac{K_{s}^{<}(j,m)}{\left(sn_{m}(s)+j~ik\right)}\left(e^{(jsn_{m}(s)+ik)x}-e^{(jsn_{m}(s)+ik)x_{m-1}}\right)\Bigg]\label{IntGretExp}\\
&+&\frac{\Phi_{s}^{<}(x)}{W_{s}}\sum_{j=+,-}j\Bigg[\frac{K_{s}^{>}(j,m)}{\left(sn_{m}(s)+j~ik\right)}\left(e^{(jsn_{m}(s)+ik)x_{m}}-e^{(jsn_{m}(s)+ik)x}\right)\nonumber\\
&+&\sum_{\alpha=m+1}^{N+1}\frac{K_{s}^{>}(j,\alpha)}{\left(sn_{\alpha}(s)+j~ik\right)}\left(e^{(jsn_{\alpha}(s)+ik)x_{\alpha}}-e^{(jsn_{\alpha}(s)+ik)x_{\alpha-1}}\right)\Bigg].\nonumber
\end{eqnarray}

Introducing this into Eq.(\ref{FieldOperatorGreenLaplaceAnnihilation}), next step is to perform the complex integration over $s$. For this purpose, the Residue theorem has to be employed considering a complex contour including the line in the complex space $l-i\Omega$, with $\Omega\in\mathbb{R}$ and closing to the left. It is important to consider that, by definition, the Laplace integration is such that the poles lie at the left of the line $l-i\Omega$. Therefore, solving the full time evolution of the contribution to the field operator of Eq.(\ref{FieldOperatorGreenLaplaceAnnihilation}) implies knowing the poles' configuration of all of the terms of Eq.(\ref{IntGretExp}). However, not all of the poles will contribute to the long-time limit ($t_{0}\rightarrow-\infty$). In fact, as it is was pointed out in Ref.\cite{TuMaRuLo}, the steady situation will be determined by the poles with zero real part, i.e., purely imaginary poles. Nevertheless, the causality property implies that the poles of the Laplace transform of the retarded Green function $\widetilde{\mathfrak{G}}_{\rm Ret}(x,x',s)$ have negative real part. The only pole with zero real part is the one at $s=0$ provided by the wronskian. Therefore, the long-time limit will be defined by the poles resulting from the spatial integration. These poles are provided by $sn_{\alpha}(s)+j~ik$ in each region.

In first place, it turns out that if a given region $\alpha$ is filled with a dissipative material ($n_{\alpha}(s)\neq1$), then $sn_{\alpha}(s)+j~ik$ has no poles with zero real part. Therefore, those terms associated to filled regions will not contribute to the steady state, i.e., it result vanishes in the long-time limit.

Thus, in principle, only regions filled with dissipationless materials may contribute to the steady situation. In particular, the vacuum regions, where $n_{\alpha}\equiv1$, and the denominator reads $s+j~ik$, providing its roots as candidate to poles ($s=-j~ik$). However, if an intermediate region of finite lenght is considered ($\alpha=2,...,N$), the corresponding term after spatial integration contains a factor $e^{(js+ik)A}-e^{(js+ik)B}$, with $A,B$ the corresponding limit of integration of the given term. This factor cancels out when $s=-j~ik$ in such a way that the limit of $(e^{(js+ik)A}-e^{(js+ik)B})/(s+j~ik)$ goes to zero (by L'Hopital rule), giving that there is no pole at $s=-j~ik$ for these terms. Clearly, this means that these terms do not contribute to the steady state.

Considering the analyzed cases at this point, the last possibility is the case where the dissipationless regions are not intermediate, having $\alpha$ equal to $1$ or $N+1$. Given the convergence of $\widetilde{\mathfrak{G}}_{\rm Ret}(x,x',s)$ for great values of $x,x'$, for $\alpha=1$ we have $K_{s}^{<}(-,1)\equiv0$, while for $\alpha=N+1$ we have $K_{s}^{>}(+,N+1)\equiv0$. This means that in these regions, outgoing waves are the only solution there. At the same time, for $\alpha=1$, given that $x_{0}=-\infty$ and that $s=l-i\Omega$ (with $l>0$), then $e^{(s+ik)x_{0}}\rightarrow0$ in the corresponding term on Eq.(\ref{IntGretExp}). Analogously, for $\alpha=N+1$, given that $x_{N+1}=+\infty$ and that $s=l-i\Omega$ (with $l>0$), we have $e^{(-s+ik)x_{N+1}}\rightarrow0$.

All in all, for dissipationless regions with $\alpha=1,N+1$, we can write the Laplace integration of Eq.(\ref{IntGretExp}) as:

\begin{eqnarray}
&&\int_{l-i\infty}^{l+i\infty}\frac{ds}{2\pi i}~e^{s(t-t_{0})}(s-i\omega_{k})\int dx'~\widetilde{\mathfrak{G}}_{\rm Ret}(x,x',s)~e^{ikx'}=\int_{l-i\infty}^{l+i\infty}\frac{ds}{2\pi i}~e^{s(t-t_{0})}(s-i\omega_{k})~\Phi_{s}^{>}(x)\frac{K_{s}^{<}(+,1)}{W_{s}}\frac{e^{(s+ik)x_{1}}}{(s+ik)}\nonumber\\
&&+\int_{l-i\infty}^{l+i\infty}\frac{ds}{2\pi i}~e^{s(t-t_{0})}(s-i\omega_{k})~\Phi_{s}^{<}(x)\frac{K_{s}^{>}(-,N+1)}{W_{s}}\frac{e^{(-s+ik)x_{N}}}{(s-ik)}+\left[\text{Terms of intermediate regions}\right].
\label{IntSIntGretExp}
\end{eqnarray}

One last simplification can be done related to the explicit calculation of the wronskian. Given that it is independent of the spatial coordinate, the wronskian can be calculated in every region, giving equalities between the different coefficients in each regions. If we calculate it in the first region ($\alpha=1$) by considering $\Phi_{s}^{<}(x')=K_{s}^{<}(+,1)~e^{sx'}$ while $\Phi_{s}^{>}(x')=e^{-sx'}+K_{s}^{>}(+,1)~e^{sx'}$ for each homogeneous solution, we obtain $W_{s}=-2s~K_{s}^{<}(+,1)$. Analogously, calculating in the last region ($\alpha=N+1$), we obtain $W_{s}=-2s~K_{s}^{>}(-,N+1)$. Then, using the first expression of the wronskian for the first term in the r.h.s. of Eq.(\ref{IntSIntGretExp}), and the second expression for the second term of the same equation, we can write:

\begin{eqnarray}
&&\int_{l-i\infty}^{l+i\infty}\frac{ds}{2\pi i}~e^{s(t-t_{0})}(s-i\omega_{k})\int dx'~\widetilde{\mathfrak{G}}_{\rm Ret}(x,x',s)~e^{ikx'}=-\int_{l-i\infty}^{l+i\infty}\frac{ds}{2\pi i}~e^{s(t-t_{0})}(s-i\omega_{k})~\Phi_{s}^{>}(x)\frac{e^{(s+ik)x_{1}}}{2s(s+ik)}\nonumber\\
&&-\int_{l-i\infty}^{l+i\infty}\frac{ds}{2\pi i}~e^{s(t-t_{0})}(s-i\omega_{k})~\Phi_{s}^{<}(x)\frac{e^{(-s+ik)x_{N}}}{2s(s-ik)}+\left[\text{Terms of intermediate regions}\right].
\label{IntSIntGretExpWronskian}
\end{eqnarray}

Taking into account that the long-time limit of this integral will be dominated by the poles with zero real part, each of the first two terms on r.h.s. of the last equation can be written as the sum of the residue at $s=0$, the corresponding at the poles $s=\pm ik$ and residue that will not contribute to the steady situation. It is worth noting that the pole at $s=0$ makes the parameter $l$ that define the Laplace antitransform to be neccesarily positive. Thus, the new poles with zero real part resulting from the spatial integration are always located at the left of the line $l+i\Omega$ and then, inside the contour of integration, as it is needed.

Now, as we mentioned before, in general (except for the first and last region) the homogeneous solutions in the region $\alpha$ are given by $\Phi_{s}^{\lessgtr}(x)=\sum_{j=+,-}K_{s}^{\lessgtr}(j,\alpha)~e^{jsn_{\alpha}(s)x}$, then each exponential contribute to different causality restrictions for each term.

The first integral splits in two, each one containing the exponential factors $e^{s(t-t_{0}+jn_{\alpha}(s)x+x_{1})}$ for each $j$. The exponents define the convergence of the integral and in which direction the contour has to be closed to use the Residue theorem. It is clear that closing to the right will give the integral equal zero due to the lack of poles inside the contour. Therefore, the integral is different from zero when the contours close to the left. To obtain the restriction in terms of $t,t_{0},x$ and $x_{1}$ resulting in each integral, as in Refs.\cite{KnollLeonhardt,Jackson}, we analyze the case of large $|s|$. In that case, $n_{\alpha}(s)\rightarrow1$ and the exponents tends to $s(t-t_{0}+jx+x_{1})$. It turns out that the contour must be closed to the left when $t-t_{0}+jx+x_{1}>0$. It should be noted that these restrictions, regardless its deduction needs large $|s|$, are valid for the entire integral, and express the natural mathematical manifestation of causality and the retardation effects that take place in the transient stage.

For the second integral the situation is analogous, and the restrictions are $t-t_{0}+jx-x_{N}>0$. It is clear that the terms of intermediate regions in Eq.(\ref{IntSIntGretExpWronskian}) also give causality restrictions but, as those terms will not contribute to the steady situation, we do not analyze them.

The generalization of this deduction for the first and last regions ($\alpha=1$ or $N+1$ respectively) is straightforward. However, for the demonstration we follow the general situation of intermediate regions.

Therefore, we have:

\begin{eqnarray}
&&\int_{l-i\infty}^{l+i\infty}\frac{ds}{2\pi i}~e^{s(t-t_{0})}(s-i\omega_{k})\int dx'~\widetilde{\mathfrak{G}}_{\rm Ret}(x,x',s)~e^{ikx'}=\label{IntSIntGretExpResidue1}\\
&&=-\sum_{j=+,-}\Theta(t-t_{0}+jx+x_{1})\Bigg({\rm Res}\left[e^{s(t-t_{0})}(s-i\omega_{k})~K_{s}^{>}(j,\alpha)~e^{jsn_{\alpha}(s)x}\frac{e^{(s+ik)x_{1}}}{2s(s+ik)},0\right]\nonumber\\
&&+~{\rm Res}\left[e^{s(t-t_{0})}(s-i\omega_{k})~K_{s}^{>}(j,\alpha)~e^{jsn_{\alpha}(s)x}\frac{e^{(s+ik)x_{1}}}{2s(s+ik)},-ik\right]\Bigg)+\left[\text{Residue with vanishing long-time limit}\right]\nonumber\\
&&-\sum_{j=+,-}\Theta(t-t_{0}+jx-x_{N})\Bigg({\rm Res}\left[e^{s(t-t_{0})}(s-i\omega_{k})~K_{s}^{<}(j,\alpha)~e^{jsn_{\alpha}(s)x}\frac{e^{(-s+ik)x_{N}}}{2s(s-ik)},0\right]\nonumber\\
&&+~{\rm Res}\left[e^{s(t-t_{0})}(s-i\omega_{k})~K_{s}^{<}(j,\alpha)~e^{jsn_{\alpha}(s)x}\frac{e^{(-s+ik)x_{N}}}{2s(s-ik)},ik\right]\Bigg)\nonumber\\
&&+\left[\text{Residue with vanishing long-time limit}\right]+\left[\text{Residue of terms of intermediate regions}\right].\nonumber
\end{eqnarray}

It turns out that a explicit calculation of the poles at $s=0$ result in time and spatial independent quantities (after assuming that $K_{0}^{<}(+,\alpha)=1=K_{0}^{>}(-,\alpha)$, while $K_{0}^{<}(-,\alpha)=0=K_{0}^{>}(+,\alpha)$ as it happens in general for the zero frequency solutions due to the scattering properties of the coefficients for every $\alpha$). The calculation of the other residue at $s=\pm ik$ is straightforward, giving:

\begin{eqnarray}
&&\int_{l-i\infty}^{l+i\infty}\frac{ds}{2\pi i}~e^{s(t-t_{0})}(s-i\omega_{k})\int dx'~\widetilde{\mathfrak{G}}_{\rm Ret}(x,x',s)~e^{ikx'}=\frac{1}{2}\frac{\omega_{k}}{k}\left[\Theta(t-t_{0}+x+x_{1})~e^{ikx_{1}}-\Theta(t-t_{0}-x-x_{N})~e^{ikx_{N}}\right]\nonumber\\
&&-\frac{1}{2}\sum_{j=+,-}\Theta(t-t_{0}+jx+x_{1})~e^{-ik(t-t_{0})}\left(1+\frac{\omega_{k}}{k}\right)~K_{-ik}^{>}(j,\alpha)~e^{-jikn_{\alpha}(-ik)x}\nonumber\\
&&-\frac{1}{2}\sum_{j=+,-}\Theta(t-t_{0}+jx-x_{N})~e^{ik(t-t_{0})}\left(1-\frac{\omega_{k}}{k}\right)~K_{ik}^{<}(j,\alpha)~e^{jikn_{\alpha}(ik)x}\label{IntSIntGretExpResidue}\\
&&+\left[\text{Residue with vanishing long-time limit}\right]+\left[\text{Residue of terms of intermediate regions}\right],\nonumber
\end{eqnarray}

\noindent where the first two terms correspond to the poles at $s=0$ of both integrals.

The crucial point is that, when considering the long-time limit ($t_{0}\rightarrow-\infty$), all the Heaviside functions present go to $1$. Thus, on one hand, the first two terms (associated to the pole at $s=0$) become time and spatial independent and, as we shall see, they will not contribute to the calculations relative to the energy-momentum tensor expectation value. On the other hand, the sums over $j$ in the third and forth terms adds up to the homogeneous solutions $\Phi^{\lessgtr}$ again. The rest terms vanish in the long-time limit. Ignoring the oscillatory dependence on $t_{0}$ of the third and forth terms and that $\frac{\omega_{k}}{k}\equiv sgn(k)$, we can write:

\begin{eqnarray}
\int_{l-i\infty}^{l+i\infty}\frac{ds}{2\pi i}~e^{s(t-t_{0})}(s-i\omega_{k})\int dx'~\widetilde{\mathfrak{G}}_{\rm Ret}(x,x',s)~e^{ikx'}&\longrightarrow&\frac{1}{2}{\rm sgn}(k)\left[e^{ikx_{1}}-e^{ikx_{N}}\right]-\frac{1}{2}~e^{-ik(t-t_{0})}\left(1+\frac{\omega_{k}}{k}\right)\Phi_{-ik}^{>}(x)\nonumber\\
&&-\frac{1}{2}~e^{ik(t-t_{0})}\left(1-\frac{\omega_{k}}{k}\right)\Phi_{ik}^{<}(x).
\end{eqnarray}

Considering that $1\pm\frac{\omega_{k}}{k}\equiv2~\Theta(\pm k)$ and assuming that the homogeneous solution satisfies $\left(\Phi_{-ik}(x)\right)^{*}=\Phi_{ik}(x)$ (or analogously, that $\Phi_{s}(x)$ is real for real $s$), inserting this into Eq.(\ref{FieldOperatorGreenLaplaceAnnihilation}) and taking the long-time limit ($t_{0}\rightarrow-\infty$) by considering all the properties of each term mentioned before, we finally obtain Eq.(\ref{LongTimeFieldOperatorIC}).

\section{Steady Volume Elements' Contribution from Complex Analysis}\label{AppC}

In this section, we follow the same approach as in Appendix \ref{AppB}, but to address the long-time limit of the field operator for the volume elements' contribution given in Eq.(\ref{LongTimeFieldOperatorA}).

Therefore, our starting point is Eq.(\ref{PhiAFull}). Writing the Green function in terms of its Laplace transform and making the substitution $\tau=t'-t_{0}$, the contribution to the field operator reads:

\begin{equation}
\widehat{\phi}_{\rm A}^{(+)}(x,t)=-\int dx'~\frac{4\pi\eta eC(x')}{\sqrt{2m\omega_{0}}}~\widehat{b}_{0,x'}(t_{0})\int_{l-i\infty}^{l+i\infty}\frac{ds}{2\pi i}~\widetilde{\mathfrak{G}}_{\rm Ret}(x,x',s)\int_{0}^{t-t_{0}}d\tau~e^{s(t-t_{0}-\tau)}\left(\dot{G}_{1}(\tau)-i\omega_{0}~\dot{G}_{2}(\tau)\right).
\end{equation}

It is worth noting that the integral over $\tau$ is a convolution. Its Laplace transform is a product of the Laplace transform of $e^{st}$ and $\dot{G}_{1}(t)-i\omega_{0}~\dot{G}_{2}(t)$. Therefore, the convolution can be written as an antitransform too, and the contribution to field operator can be written as:

\begin{equation}
\widehat{\phi}_{\rm A}^{(+)}(x,t)=-\int dx'~\frac{4\pi\eta eC(x')}{\sqrt{2m\omega_{0}}}~\widehat{b}_{0,x'}(t_{0})\int_{l-i\infty}^{l+i\infty}\frac{ds}{2\pi i}~\widetilde{\mathfrak{G}}_{\rm Ret}(x,x',s)\int_{l'-i\infty}^{l'+i\infty}\frac{dz}{2\pi i}\frac{e^{z(t-t_{0})}}{(z-s)}\left(z(z-i\omega_{0})\widetilde{G}_{2}(z)-1\right),
\end{equation}

\noindent where $l'>l$ to have a well-defined Laplace transform of $e^{st}$, ensuring the convergence of the integral and the causality property at once.

This expression allow us to write the convolution in terms of the residue of the poles of the integrand. Due to the causality property, $\widetilde{G}_{2}$ has poles with non-positive real parts. Then, as $l'>l$, the integral over $z$ can be written as the residuum at $z=s$ (associated to $1/(z-s)$) plus all the residue with non-positive real parts that will give terms that vanish at the long-time limit ($t_{0}\rightarrow-\infty$). Thus, we obtain:

\begin{eqnarray}
\widehat{\phi}_{\rm A}^{(+)}(x,t)&=&-\int dx'~\frac{4\pi\eta eC(x')}{\sqrt{2m\omega_{0}}}~\widehat{b}_{0,x'}(t_{0})\int_{l-i\infty}^{l+i\infty}\frac{ds}{2\pi i}~e^{s(t-t_{0})}~\widetilde{\mathfrak{G}}_{\rm Ret}(x,x',s)\left(s(s-i\omega_{0})\widetilde{G}_{2}(s)-1\right)\nonumber\\
&&+\left[\text{Terms with vanishing long-time limit}\right].
\end{eqnarray}

Again, the causality of the retarded Green functions $\mathfrak{G}_{\rm Ret},G_{2}$ gives that the integrand in the integral over $s$ have poles with non-positive real parts. The only pole having zero real part is the one at $s=0$ provided by the wronskian contained in $\widetilde{\mathfrak{G}}_{\rm Ret}$, which is the only one that will not vanish at the long-time limit. It should be noted that only the second term of $s(s-i\omega_{0})\widetilde{G}_{2}(s)-1$ will have $s=0$ as a pole since the for the first term the denominator that provides the pole cancels with $s$ on the numerator.

Considering the expression of the Green function of Eq.(\ref{GreenFunctionCollin}) and the general form for the homogeneous solutions $\Phi_{s}^{\lessgtr}$, the Green function can be written as:

\begin{equation}
\mathfrak{G}_{\rm Ret}(x,x',s)=\frac{1}{W_{s}}\sum_{j,k=+,-}\left[\Theta(x-x')~K_{s}^{<}(j,\alpha)~K_{s}^{>}(k,\beta)+\Theta(x'-x)~K_{s}^{<}(k,\beta)~K_{s}^{>}(j,\alpha)\right]e^{s\left(jn_{\alpha}(s)x'+kn_{\beta}(s)x\right)},
\label{FieldRetGreenLaplaceDoubleSum}
\end{equation}

\noindent for arbitrary $x$ in the $\alpha-$region and $x'$ in the $\beta-$region.

Now, by taking in account the properties of the coefficients $K_{s}^{\lessgtr}$ at $s=0$ and the analysis associated to the causality of the transient stage, we can calculate the residuum at $s=0$, obtaining for every $x$:

\begin{eqnarray}
\widehat{\phi}_{\rm A}^{(+)}(x,t)&=&-\int dx'~\frac{4\pi\eta eC(x')}{\sqrt{2m\omega_{0}}}~\widehat{b}_{0,x'}(t_{0})~\frac{\Theta(t-t_{0}+x'-x)~\Theta(t-t_{0}-x'+x)}{2}\nonumber\\
&&+\left[\text{Terms with vanishing long-time limit}\right].
\end{eqnarray}

Thus, taking the long-time limit ($t_{0}\rightarrow-\infty$), we easily obtain Eq.(\ref{LongTimeFieldOperatorA}).

\section{Steady Baths' Contribution from Complex Analysis}\label{AppD}

The present section shows how Eq.(\ref{LongTimeFieldOperatorB}) can be obtained taking Eq.(\ref{PhiBFull}) as the starting point. The approach will be the same as in the last two sections for the others two contribution, by separating the steady contribution by analyzing the poles configuration.

First step consists in writing the field's retarded Green function as a Laplace antitransform, re-writing Eq.(\ref{PhiBFull}) as:

\begin{equation}
\widehat{\phi}_{\rm B}(x,t)=-\int dx'~4\pi\eta eC(x')\int_{l-i\infty}^{l+i\infty}\frac{ds}{2\pi i}~\widetilde{\mathfrak{G}}_{\rm Ret}(x,x',s)\int_{t_{0}}^{t}dt'~e^{s(t-t')}\int_{t_{0}}^{t'}~d\tau~\dot{G}_{2}(t'-\tau)~\frac{\widehat{F}_{x'}(\tau-t_{0})}{m}.
\end{equation}

The resulting integrals over $t'$ and $\tau$ are basically a double convolution and we can write it in terms of its Laplace transform:

\begin{equation}
\int_{t_{0}}^{t}dt'~e^{s(t-t')}\int_{t_{0}}^{t'}~d\tau~\dot{G}_{2}(t'-\tau)~\frac{\widehat{F}_{x'}(\tau-t_{0})}{m}=\int_{l'-i\infty}^{l'+i\infty}\frac{dz}{2\pi i}~e^{z(t-t_{0})}\frac{z}{(z-s)}~\widetilde{G}_{2}(z)\frac{\widehat{\widetilde{F}}_{x'}(z)}{m}.
\end{equation}

\noindent where, as for the volume elements' contribution, we have to take $l'>l$ in order to verify the convergence requirement of the function $e^{st}$. Given this, we can re-write the double convolution in terms of its residue associated to the poles of the integrand in the r.h.s. of the last expression. These are the pole at $z=s$, the poles provided by the Laplace transform of the stochastic force operator and the poles associated to $\widetilde{G}_{2}$, which gives terms that goes to zero in the long-time limit:

\begin{eqnarray}
\int_{t_{0}}^{t}dt'~e^{s(t-t')}\int_{t_{0}}^{t'}~d\tau~\dot{G}_{2}(t'-\tau)~\frac{\widehat{F}_{x'}(\tau-t_{0})}{m}&=&e^{s(t-t_{0})}~s~\widetilde{G}_{2}(s)~\frac{\widehat{\widetilde{F}}_{x'}(s)}{m}+\left[\text{Residue of the poles of $\widehat{\widetilde{F}}_{x'}$}\right]\nonumber\\
&+&\left[\text{Terms with vanishing long-time limit}\right].
\label{DoubleConvolutionResidue}
\end{eqnarray}

Now, by considering Eq.(\ref{QBMOperadorFdet}), the Laplace transform of the stochastic force operator is given by:

\begin{equation}
\widehat{\widetilde{F}}_{x'}(z)=\sum_{n}\frac{\lambda_{n}}{\sqrt{2m_{n}\omega_{n}}}\left(\frac{1}{(z+i\omega_{n})}~\widehat{b}_{n,x'}(t_{0})+\frac{1}{(z-i\omega_{n})}~\widehat{b}_{n,x'}^{\dag}(t_{0})\right),
\label{QBMOperadorFLaplace}
\end{equation}

\noindent which presents poles at $z=\pm i\omega_{n}$ for each term, and therefore the second term of the r.h.s. of Eq.(\ref{DoubleConvolutionResidue}) can be written as:

\begin{eqnarray}
\left[\text{Residue of the poles of $\widehat{\widetilde{F}}_{x'}$}\right]&=&{\rm Res}\left[e^{z(t-t_{0})}\frac{z~\widetilde{G}_{2}(z)}{(z-s)}\frac{\widehat{\widetilde{F}}_{x'}(z)}{m},-i\omega_{n}\right]+{\rm Res}\left[e^{z(t-t_{0})}\frac{z~\widetilde{G}_{2}(z)}{(z-s)}\frac{\widehat{\widetilde{F}}_{x'}(z)}{m},i\omega_{n}\right]\\
&=&\frac{1}{m}\sum_{n}\frac{\lambda_{n}~i\omega_{n}}{\sqrt{2m_{n}\omega_{n}}}\left[\frac{\widetilde{G}_{2}(-i\omega_{n})}{(s+i\omega_{n})}~e^{-i\omega_{n}(t-t_{0})}~\widehat{b}_{n,x'}(t_{0})-\frac{\widetilde{G}_{2}(i\omega_{n})}{(s-i\omega_{n})}~e^{i\omega_{n}(t-t_{0})}~\widehat{b}_{n,x'}^{\dag}(t_{0})\right].\nonumber
\end{eqnarray}

At this point, we can write the last expression as an integral over a continuous frequency $\omega$ by introducing Dirac delta functions:

\begin{eqnarray}
\left[\text{Residue of the poles of $\widehat{\widetilde{F}}_{x'}$}\right]&=&\int_{-\infty}^{+\infty}\frac{d\omega}{2\pi}~e^{-i\omega t}\frac{i\omega~\widetilde{G}_{2}(-i\omega)}{m(s+i\omega)}~2\pi\sum_{n}\frac{\lambda_{n}}{\sqrt{2m_{n}\omega_{n}}}\Big[\delta(\omega-\omega_{n})~e^{i\omega_{n}t_{0}}\widehat{b}_{n,x'}(t_{0})+\delta(\omega+\omega_{n})\nonumber\\
&\times&~e^{-i\omega_{n}t_{0}}\widehat{b}_{n,x'}^{\dag}(t_{0})\Big]\nonumber\\
&=&\int_{-\infty}^{+\infty}\frac{d\omega}{2\pi}~e^{-i\omega t}\frac{i\omega~\overline{G}_{2}(\omega)}{(s+i\omega)}\frac{\widehat{\overline{F}}_{x'}(\omega)}{m},
\end{eqnarray}

\noindent where we have used that $\widetilde{f}(-i\omega)=\overline{f}(\omega)$ for every causal function as the retarded Green function $G_{2}$, being $\overline{f}(\omega)$ the Fourier transform of the given causal function $f(t)$ (which also admits a Laplace transform). Moreover, $\widehat{\overline{F}}_{x'}(\omega)$ turns out to be the Fourier transform of the stochastic force operator, considering the time variable in Eq.(\ref{QBMOperadorFdet}) extended to all the real domain.

Considering the last expression and replacing Eq.(\ref{DoubleConvolutionResidue}) into the field's contribution, we have:

\begin{eqnarray}
\widehat{\phi}_{\rm B}(x,t)&=&-\int dx'~4\pi\eta eC(x')\int_{l-i\infty}^{l+i\infty}\frac{ds}{2\pi i}~e^{s(t-t_{0})}~s~\widetilde{G}_{2}(s)~\widetilde{\mathfrak{G}}_{\rm Ret}(x,x',s)~\frac{\widehat{\widetilde{F}}_{x'}(s)}{m}\\
&-&\int dx'~4\pi\eta eC(x')\int_{-\infty}^{+\infty}\frac{d\omega}{2\pi}~e^{-i\omega t}~i\omega~\overline{G}_{2}(\omega)~\frac{\widehat{\overline{F}}_{x'}(\omega)}{m}\int_{l-i\infty}^{l+i\infty}\frac{ds}{2\pi i}\frac{\widetilde{\mathfrak{G}}_{\rm Ret}(x,x',s)}{(s+i\omega)}\nonumber\\
&+&\left[\text{Terms with vanishing long-time limit}\right]\nonumber.
\end{eqnarray}

Nevertheless, the integral over $s$ of the second term of the last equation can be worked out further. Considering it as a limit of a Laplace antitransform:

\begin{equation}
\int_{l-i\infty}^{l+i\infty}\frac{ds}{2\pi i}\frac{\widetilde{\mathfrak{G}}_{\rm Ret}(x,x',s)}{(s+i\omega)}=\lim_{t'\rightarrow t_{0}}\int_{l-i\infty}^{l+i\infty}\frac{ds}{2\pi i}~e^{s(t'-t_{0})}~\frac{\widetilde{\mathfrak{G}}_{\rm Ret}(x,x',s)}{(s+i\omega)},
\end{equation}

\noindent where the r.h.s. result to be the antitransform of a convolution of the functions associated to the transforms $\widetilde{\mathfrak{G}}_{\rm Ret}(x,x',s)$ and $1/(s+i\omega)$. Therefore, we can write the convolution and take the limit on the integral, which clearly vanishes due to the fact that both integration limits become the same:

\begin{equation}
\int_{l-i\infty}^{l+i\infty}\frac{ds}{2\pi i}\frac{\widetilde{\mathfrak{G}}_{\rm Ret}(x,x',s)}{(s+i\omega)}=\lim_{t'\rightarrow t_{0}}\int_{t_{0}}^{t'}d\tau~\mathfrak{G}_{\rm Ret}(x,x',t'-\tau)~e^{-i\omega(\tau-t_{0})}\equiv 0.
\end{equation}

Hence, the contribution of the baths to the field operator reads:

\begin{eqnarray}
\widehat{\phi}_{\rm B}(x,t)&=&-\int dx'~4\pi\eta eC(x')\int_{l-i\infty}^{l+i\infty}\frac{ds}{2\pi i}~e^{s(t-t_{0})}~s~\widetilde{G}_{2}(s)~\widetilde{\mathfrak{G}}_{\rm Ret}(x,x',s)~\frac{\widehat{\widetilde{F}}_{x'}(s)}{m}\\
&+&\left[\text{Terms with vanishing long-time limit}\right]\nonumber.
\end{eqnarray}

Again, for the Laplace antitransform of the first term in r.h.s., we can proceed as before, and write it in terms of the residue of the integrand. This time, we have the poles of $\widehat{\widetilde{F}}_{x'}$ (located at $s=\pm i\omega_{n}$) and the ones with negative real part provided by $\widetilde{G}_{2}$ and $\widetilde{\mathfrak{G}}_{\rm Ret}$, but excluding the pole at $s=0$ due to the factor $s$ present on the integrand which prevents it. As we mentioned before, the poles with negative real part will result in terms that vanish at the long-time limit. 



We can proceed as before to write the terms in the last expression as an integral over $\omega$, finally obtaining for the contribution of the baths to the field operator:

\begin{eqnarray}
\widehat{\phi}_{\rm B}(x,t)&=&-\int dx'~\frac{4\pi\eta eC(x')}{m}\int_{-\infty}^{+\infty}\frac{d\omega}{2\pi}~e^{-i\omega t}~(-i\omega)~\frac{\overline{G}_{2}(\omega)}{W_{-i\omega}}\sum_{j,k=+,-}\Theta\left(t-t_{0}+jx'+kx\right)\Big(\Theta(x-x')~K_{-i\omega}^{<}(j,\alpha)\nonumber\\
&\times&~K_{-i\omega}^{>}(k,\beta)+\Theta(x'-x)~K_{-i\omega}^{<}(k,\beta)~K_{-i\omega}^{>}(j,\alpha)\Big)~e^{-i\omega\left(jn_{\alpha}(-i\omega)x'+kn_{\beta}(-i\omega)x\right)}~\widehat{\overline{F}}_{x'}(\omega)\\
&+&\left[\text{Terms with vanishing long-time limit}\right]\nonumber.
\end{eqnarray}

Taking the long-time limit of this expression is straightforward to finally obtain Eq.(\ref{LongTimeFieldOperatorB}).

\section{Green Function for Two Finite Width Plates Configuration}\label{AppE}

To calculate each contribution to the total force through Eqs.(\ref{TMuNuB}) and (\ref{TMuNuFdeK}), we need the modified modes $\Phi_{s}^{\lessgtr}$ (and consequently, the Green function) obtained from Eq.(\ref{EcMovGreenLAPLACE}) for the specific problem. This section is devoted to give those expressions.

For the present problem of two different slabs of width $d$ separated by a distance $a$, the modified modes are given by:

\begin{eqnarray}
\Phi^{<}_{s}(x)= \left\{
\begin{array}{lr rl}
T_{s}^{<}~e^{sx}, &&& \text{for}~x<-d-\frac{a}{2}\\
E_{s}^{<}~e^{sn_{L}x}+F_{s}^{<}~e^{-sn_{L}x}, &&& \text{for}~-d-\frac{a}{2}<x<-\frac{a}{2}\\
C_{s}^{<}~e^{sx}+D_{s}^{<}~e^{-sx}, &&& \text{for}~-\frac{a}{2}<x<\frac{a}{2}\\
A_{s}^{<}~e^{sn_{R}x}+B_{s}^{<}~e^{-sn_{R}x}, &&& \text{for}~\frac{a}{2}<x<d+\frac{a}{2}\\
e^{sx}+R_{s}^{<}~e^{-sx}, &&& \text{for}~d+\frac{a}{2}<x\\
\end{array}
\right.
\end{eqnarray}

\begin{eqnarray}
\Phi^{>}_{s}(x)= \left\{
\begin{array}{lr rl}
e^{-sx}+R_{s}^{>}~e^{sx}, &&& \text{for}~x<-d-\frac{a}{2}\\
A_{s}^{>}~e^{-sn_{L}x}+B_{s}^{>}~e^{sn_{L}x}, &&& \text{for}~-d-\frac{a}{2}<x<-\frac{a}{2}\\
C_{s}^{>}~e^{-sx}+D_{s}^{>}~e^{sx}, &&& \text{for}~-\frac{a}{2}<x<\frac{a}{2}\\
E_{s}^{>}~e^{-sn_{R}x}+F_{s}^{>}~e^{sn_{R}x}, &&& \text{for}~\frac{a}{2}<x<d+\frac{a}{2}\\
T_{s}^{>}~e^{-sx}, &&& \text{for}~d+\frac{a}{2}<x\\
\end{array}
\right.
\end{eqnarray}

\noindent where the coefficients for each homogeneous solution can be found in Ref.\cite{Dorota1993}:

\begin{equation}
R_{s}^{>}=\left[r_{L}+\frac{r_{R}t_{L}^{2}~e^{-2sa}}{1-r_{L}r_{R}~e^{-2sa}}\right]e^{s(a+2d)},~~~~~T_{s}^{>}=\frac{t_{R}t_{L}~e^{2sd}}{1-r_{L}r_{R}~e^{-2sa}},
\label{RandTCoefficients}
\end{equation}

\begin{equation}
C_{s}^{>}=e^{-sd}~\frac{T_{s}^{>}}{t_{R}}
,~~~~~D_{s}^{>}=e^{-s(a+d)}~\frac{r_{R}}{t_{R}}~T_{s}^{>}
,
\label{CandDCoefficients}
\end{equation}

\begin{equation}
A_{s}^{>}=\frac{(n_{L}+1)}{2n_{L}}e^{s(1-n_{L})\left(\frac{a}{2}+d\right)}\left[1-r_{n_{L}}~R_{s}^{>}~e^{-s(a+2d)}\right],~~~~~B_{s}^{>}=\frac{(n_{L}+1)}{2n_{L}}e^{s(1+n_{L})\left(\frac{a}{2}+d\right)}\left[R_{s}^{>}~e^{-s(a+2d)}-r_{n_{L}}\right],
\label{AandBCoefficients}
\end{equation}

\begin{equation}
E_{s}^{>}=\frac{(n_{R}+1)}{2n_{R}}e^{s(n_{R}-1)\left(\frac{a}{2}+d\right)}~T_{s}^{>},~~~~~F_{s}^{>}=\frac{(n_{R}-1)}{2n_{R}}e^{-s(n_{R}+1)\left(\frac{a}{2}+d\right)}~T_{s}^{>},
\label{EandFCoefficients}
\end{equation}

\noindent where we have given all the coefficients in terms of the reflection and transmission coefficients of the two plates configuration ($R_{s}^{>}$ and $T_{s}^{>}$). Moreover, $r_{L,R}$ and $t_{L,R}$ are the reflection and transmission coefficients for the left and right plates respectively:

\begin{equation}
r_{i}=\frac{r_{n_{i}}\left(1-e^{-2sn_{i}d}\right)}{\left(1-r_{n_{i}}^{2}e^{-2sn_{i}d}\right)},~~~~~t_{i}=\frac{4n_{i}}{(n_{i}+1)^{2}}\frac{e^{-sn_{i}d}}{\left(1-r_{n_{i}}^{2}e^{-2sn_{i}d}\right)},
\label{OnePlateRandTCoefficients}
\end{equation}

\noindent with $r_{n_{i}}=\frac{1-n_{i}}{1+n_{i}}$, the reflection coefficient of a surface of refractive index $n_{i}$.

It should be noted that the $<-$coefficients are obtained from the given ones by the interchange of $L$ and $R$ in the expressions. Considering this, it turns out that $T_{s}^{>}=T_{s}^{<}$, so the superscript for this coefficient can be omitted.

For the given configuration of finite width plates, the boundary conditions on the modes were continuity of the mode and its spatial derivative at the interfaces between the material slabs and the surrounding vacuum.

On the other hand, to calculate the contribution of the baths to the Casimir force, we also need the Green function (or its Laplace transform) with the field point $x$ in the regions outside and between the plates.

Therefore, taking $x<-d-\frac{a}{2}$, through Eq.(\ref{GreenFunctionCollin}), the Laplace transform of the retarded Green function reads:

\begin{eqnarray}
\widetilde{\mathfrak{G}}_{\rm Ret}(x,x',s)=-\frac{1}{2s}\left\{
\begin{array}{lr rl}
e^{sx'}\left(e^{-sx}+R_{s}^{>}~e^{sx}\right), &&& \text{for}~x'<x<-d-\frac{a}{2}\\
(e^{-sx'}+R_{s}^{>}~e^{sx'})e^{sx}, &&& \text{for}~x<x'<-d-\frac{a}{2}\\
(A_{s}^{>}~e^{-sn_{L}x'}+B_{s}^{>}~e^{sn_{L}x'})e^{sx}, &&& \text{for}~x<-d-\frac{a}{2}<x'<-\frac{a}{2}\\
(C_{s}^{>}~e^{-sx'}+D_{s}^{>}~e^{sx'})e^{sx}, &&& \text{for}~x<-d-\frac{a}{2}<-\frac{a}{2}<x'<\frac{a}{2}\\
(E_{s}^{>}~e^{-sn_{R}x'}+F_{s}^{>}~e^{sn_{R}x'})e^{sx}, &&& \text{for}~x<-d-\frac{a}{2}<\frac{a}{2}<x'<d+\frac{a}{2}\\
T_{s}~e^{-sx'}~e^{sx}, &&& \text{for}~x<-d-\frac{a}{2}<d+\frac{a}{2}<x'\\
\end{array}
\right.
\end{eqnarray}

For the case where $-\frac{a}{2}<x<\frac{a}{2}$, we have:

\begin{eqnarray}
\widetilde{\mathfrak{G}}_{\rm Ret}(x,x',s)=-\frac{1}{2s}\left\{
\begin{array}{lr rl}
T_{s}~e^{sx'}\frac{e^{-sd}}{t_{R}}\left(e^{-sx}+r_{R}~e^{-sa}~e^{sx}\right), &&& \text{for}~x'<-d-\frac{a}{2}<-\frac{a}{2}<x<\frac{a}{2}\\
(E_{s}^{<}e^{sn_{L}x'}+F_{s}^{<}e^{-sn_{L}x'})\frac{e^{-sd}}{t_{R}}\left(e^{-sx}+r_{R}e^{-sa}e^{sx}\right), &&& \text{for}~-d-\frac{a}{2}<x'<-\frac{a}{2}<x<\frac{a}{2}\\
(C_{s}^{<}~e^{sx'}+D_{s}^{<}~e^{-sx'})\frac{e^{-sd}}{t_{R}}\left(e^{-sx}+r_{R}~e^{-sa}~e^{sx}\right), &&& \text{for}~-\frac{a}{2}<x'<x<\frac{a}{2}\\
(C_{s}^{>}~e^{-sx'}+D_{s}^{>}~e^{sx'})\frac{e^{-sd}}{t_{L}}\left(e^{sx}+r_{L}~e^{-sa}~e^{-sx}\right), &&& \text{for}~-\frac{a}{2}<x<x'<\frac{a}{2}\\
(E_{s}^{>}e^{-sn_{R}x'}+F_{s}^{>}e^{sn_{R}x'})\frac{e^{-sd}}{t_{L}}\left(e^{sx}+r_{L}e^{-sa}e^{-sx}\right), &&& \text{for}~-\frac{a}{2}<x<\frac{a}{2}<x'<d+\frac{a}{2}\\
T_{s}~e^{-sx'}\frac{e^{-sd}}{t_{L}}\left(e^{sx}+r_{L}~e^{-sa}~e^{-sx}\right), &&& \text{for}~-\frac{a}{2}<x<\frac{a}{2}<d+\frac{a}{2}<x'.\\
\end{array}
\right.
\end{eqnarray}

\section{Limit Cases: Dissipationless Material and the Lifshitz Formula}\label{AppF}

In this section, we show how the general expressions for the force given in Eqs.(\ref{CasimirForceICFull}), (\ref{CasimirForceBathFull}) and (\ref{CasimirForceFull}) allow us to recover two well-known results for the Casimir force. One of the limit cases is to obtain the Casimir force existing between two slabs of arbitrary thickness made of materials without dissipation. The other one is how to obtain the force between two half-spaces (infinite width) of dissipative materials at thermal equilibrium, which is known as the Lifshitz formula.

\subsection{Material Without Dissipation}

The first limit-case to verify is the Casimir force between two plates of finite width and different materials without dissipation. Considering the definition of the refractive index in each point of the material, given by the Eq.(\ref{RefractionIndexX}), setting dissipation equal to zero implies taking $\overline{\gamma}_{\rm L,R}(\omega)\equiv 0$, i.e., damping kernel equal to zero for each plate. Therefore, the refractive indexes (and consequently the permittivity function) simplifies to:

\begin{equation}
n_{i}^{2}(-i\omega)=\varepsilon_{i}(\omega)=1+\frac{\omega_{{\rm Pl},i}^{2}}{(\omega_{0,i}^{2}-\omega^{2})},
\end{equation}

\noindent which is a real function of $\omega$.

However, real frequency-dependent permittivities violate Kramers-Kronig relations and, consequently, are forbidden in order to respect physical causality.

Hence, setting dissipation equal to zero is not enough to obtain a well-defined physical model without dissipation. To finally obtain the correct model, as it done in Ref.\cite{CTPScalar}, we have to take the zeroth order of this last expression by setting $\omega=0$. In this case, we obtain:

\begin{equation}
\varepsilon_{i,\rm ND}=1+\frac{\omega_{{\rm Pl},i}^{2}}{\omega_{0,i}^{2}},
\end{equation}

\noindent which is a real but frequency-independent function. These permittivities verifies Kramers-Kronig relation trivially, since a null imaginary part implies a frequency-independent real part (see Ref.\cite{Jackson}).

All in all, taking the limit-case of materials without dissipation is achieved by putting $\overline{\gamma}\equiv 0$ plus take the zeroth order of the resulting permittivity.

Once we have considered how to take the limit of materials without dissipation, in any case, we have ${\rm Im}(n_{i})\equiv 0$. This also implies that $(\pm 1)(e^{\pm2\omega{\rm Im}(n_{i})d}-1)\equiv 0$. Therefore, the direct consequence is that the baths' contribution trivially vanishes:

\begin{equation}
F_{\rm C}^{\rm B}\left[a,d,\beta_{\rm B,L},\beta_{\rm B,R}\right]\Big|_{\text{No Diss}}\equiv 0.
\end{equation}

Nevertheless, the contribution of the initial conditions to the Casimir force $F_{\rm C}^{\rm IC}\left[a,d,\mathfrak{F}\right]$ does not vanishes. In fact, in this case, it is responsible of the totally of the Casimir force.

Moreover, once the dissipation is suppressed, both identities $\left|R_{-ik}^{>}\right|^{2}+\left|T_{-ik}\right|^{2}=1$ and $|r_{i}|^{2}+|t_{i}|^{2}=1$ are also recovered.

Accordingly, the total Casimir force reads:

\begin{eqnarray}
F_{\rm C}\left[a,d,\mathfrak{F}\right]\Big|_{\text{No Diss}}&\equiv &F_{\rm C}^{\rm IC}\left[a,d,\mathfrak{F}\right]\Big|_{\text{No Diss}}\nonumber\\
&=&\int_{0}^{+\infty}dk~k~\mathfrak{F}(k)\left[2-\left|C_{-ik}^{>}\right|^{2}-\left|D_{-ik}^{>}\right|^{2}-\left|C_{-ik}^{<}\right|^{2}-\left|D_{-ik}^{<}\right|^{2}\right]\nonumber\\
&=&\int_{0}^{+\infty}dk~k~\mathfrak{F}(k)\left[2-\frac{[|t_{\rm L}|^{2}(1+|r_{\rm R}|^{2})+|t_{\rm R}|^{2}(1+|r_{\rm L}|^{2})]}{|1-r_{\rm L}r_{\rm R}~e^{i2\omega a}|^{2}}\right].
\label{CasimirForceNODiss}
\end{eqnarray}

For the case of considering the same material for both plates, we have that $r_{\rm L}=r_{\rm R}=r$ and $t_{\rm L}=t_{\rm R}=t$, and the last expression straightforward simplifies to:

\begin{eqnarray}
F_{\rm C}\left[a,d,\mathfrak{F}\right]\Big|_{\text{No Diss}}^{\text{1-Mat}}=2\int_{0}^{+\infty}dk~k~\mathfrak{F}(k)\left[1-\frac{1-|r|^{4}}{|1-r^{2}~e^{i2\omega a}|^{2}}\right],
\end{eqnarray}

\noindent which is the Casimir force between two finite plates of finite width and the same material without dissipation. If we consider an equilibrium scenario, $\mathfrak{F}(k)=\coth\left(\frac{\beta_{\phi}\omega_{k}}{2}\right)$, and the result is in agreement with the result found in Ref.\cite{Dorota1990} for a formally equivalent situation.

Moreover, from this point of view for the thermal case, Eq.(\ref{CasimirForceNODiss}) is the generalization of the result of Ref.\cite{Dorota1990} for the case of plates of different materials without dissipation. Nevertheless, the result obtained is an equilibrium situation ensured by the equilibrium state for the field. Considering the result obtained for this limit case, the calculation of the Casimir force between material bodies without dissipation is formally achieved through a quantum field described in a Hilbert space that here we can match it to the Hilbert space of a free field. This is in fact what it is done in well-known Literature to obtain the Casimir force between dielectric plates without dissipation and in equilibrium (see Refs.\cite{Milonni,BordagMohideenMostepanenko}). In previous works, the result including dissipative materials is often obtained as an extension of this simplified scenario through letting the real and frequency-independent permittivity in this case to be replaced by a complex frequency-dependent function in the final result. This extension procedure in fact works in equilibrium situations because the results including and disregarding dissipation are formally the same. However, as we have seen before, non-equilibrium scenarios critically requires the introduction of degrees of freedom for the materials bodies since the extension procedure loses physical sense in successfully introducing non-equilibrium features through a particular state for the field.

\subsection{Lifshitz Formula}

The other important limit-case is the Lifshitz formula. Starting from Eqs.(\ref{CasimirForceICFull}) and (\ref{CasimirForceBathFull}), the formal procedure to recover the well-known Lifshitz's result consists on taking the limit of infinite width ($d\rightarrow+\infty$) in the expressions and then set thermal equilibrium between the baths and the field at the initial time (therefore, for the initial conditions' contribution, we will be considering $\mathfrak{F}(k)=\coth\left(\frac{\beta_{\phi}\omega_{k}}{2}\right)$ in all this section).

To successfully take the limit of infinite width on the contributions to the total force, from Eq.(\ref{OnePlateRandTCoefficients}), we have that $r_{i}\rightarrow r_{n_{i}}$, $t_{i}\rightarrow 0$ but $|t_{i}|^{2}e^{2\omega{\rm Im}(n_{i})d}\rightarrow\frac{16|n_{i}|^{2}}{|n_{i}+1|^{4}}$. Therefore, considering the definition for the different coefficients, given in Eqs.(\ref{RandTCoefficients}) and (\ref{CandDCoefficients}), and that $1-|r_{n_{i}}|^{2}=\frac{2{\rm Re}(n_{i})}{|n_{i}+1|^{2}}$, each contribution to the force reads:

\begin{equation}
F_{\rm C}^{\rm IC}\left[a,d\rightarrow+\infty,\beta_{\phi}\right]=\int_{0}^{+\infty} dk~k~\coth\left(\frac{\beta_{\phi}k}{2}\right)\left[1+|r_{n_{\rm L}}|^{2}\right],
\end{equation}

\begin{eqnarray}
F_{\rm C}^{\rm B}\left[a,d\rightarrow+\infty,\beta_{\rm B,L},\beta_{\rm B,R}\right]=\int_{0}^{+\infty}d\omega~\omega&\Bigg[&\coth\left(\frac{\beta_{\rm B,L}\omega}{2}\right)\left[1-|r_{n_{\rm L}}|^{2}\right]\left(1-\frac{\left[1+|r_{n_{\rm R}}|^{2}\right]}{|1-r_{n_{\rm L}}r_{n_{\rm R}}~e^{i2\omega a}|^{2}}\right)\nonumber\\
&-&\coth\left(\frac{\beta_{\rm B,R}\omega}{2}\right)\frac{\left[1-|r_{n_{\rm R}}|^{2}\right]\left[1+|r_{n_{\rm L}}|^{2}\right]}{|1-r_{n_{\rm L}}r_{n_{\rm R}}~e^{i2\omega a}|^{2}}\Bigg].
\end{eqnarray}

Setting thermal equilibrium between the baths and the field ($\beta_{\rm B,L}=\beta_{\rm B,R}=\beta_{\phi}\equiv\beta$), the total force reads:

\begin{eqnarray}
F_{\rm C}\left[a,d\rightarrow+\infty,\beta,\beta,\beta\right]&=&F_{\rm C}^{\rm IC}\left[a,d\rightarrow+\infty,\beta\right]+F_{\rm C}^{\rm B}\left[a,d\rightarrow+\infty,\beta,\beta\right]\nonumber\\
&=&4\int_{0}^{+\infty}dk~k~\coth\left(\frac{\beta k}{2}\right)\frac{\left[|r_{n_{\rm R}}|^{2}|r_{n_{\rm L}}|^{2}-{\rm Re}\left(r_{n_{\rm L}}r_{n_{\rm R}}~e^{i2\omega a}\right)\right]}{|1-r_{n_{\rm L}}r_{n_{\rm R}}~e^{i2\omega a}|^{2}}\nonumber\\
&=&-\int_{-\infty}^{+\infty}dk~k~\coth\left(\frac{\beta k}{2}\right){\rm Re}\left[\frac{r_{n_{\rm L}}(-ik)r_{n_{\rm R}}(-ik)~e^{i2ka}}{1-r_{n_{\rm L}}(-ik)r_{n_{\rm R}}(-ik)~e^{i2ka}}\right],
\end{eqnarray}

\noindent where we have used the fact that the integrand is even and stressed the explicit dependence on $k$ related to the definition of the coefficients given in Eq.(\ref{OnePlateRandTCoefficients}).

In order to calculate the integral through the Residue Theorem, we can consider the expansion of the thermal factor in terms of the Matsubara poles:

\begin{equation}
\coth\left(\frac{\beta k}{2}\right)=\frac{2}{\beta}\left[\frac{1}{k}+\sum_{l=1}^{+\infty}\left(\frac{1}{k+i\xi_{l}}+\frac{1}{k-i\xi_{l}}\right)\right]
\end{equation}

\noindent with $\xi_{l}=\frac{2\pi l}{\beta}$ the Matsubara frequencies.

Assuming that $k~{\rm Re}\left[\frac{r_{n_{\rm L}}r_{n_{\rm R}}e^{i2\omega a}}{1-r_{n_{\rm L}}r_{n_{\rm R}}e^{i2\omega a}}\right]$ has no poles on the upper half of the complex plane, Cauchy Theorem can be applied. By considering a contour closing in the upper half, the only poles that contribute to the integral are $i\xi_{l}$ (the denominator $1/k$ for the first term of the last equation is cancelled, giving no pole at $k=0$). Therefore, we obtain:

\begin{equation}
F_{\rm C}\left[a,d\rightarrow+\infty,\beta,\beta,\beta\right]=\frac{4\pi}{\beta}\sum_{l=1}^{+\infty}\xi_{l}\frac{r_{n_{\rm L}}(\xi_{l})r_{n_{\rm R}}(\xi_{l})~e^{-2\xi_{l}a}}{(1-r_{n_{\rm L}}(\xi_{l})r_{n_{\rm R}}(\xi_{l})~e^{-2\xi_{l}a)})},
\end{equation}

\noindent where from Eq.(\ref{OnePlateRandTCoefficients}) we have that $r_{n_{i}}(s)$ is real for real $s$.

The last equation corresponds exactly to the finite-temperature Lifshitz formula for two plates of different dissipative materials (being the generalization of the result found in Ref.\cite{LombiMazziRL}).

\end{document}